\documentclass[12pt]{iopart}

\usepackage{graphicx}% Include figure files
\usepackage{dcolumn}% Align table columns on decimal point
\usepackage{bm}% bold math
\usepackage{bbm}
\usepackage{epsfig}
\usepackage{mathrsfs}
\usepackage{stmaryrd}
\usepackage{color}
\usepackage{dsfont}
\usepackage{iopams}
\usepackage{subeqn}
\newcommand{\bra}[1]{\langle\,{#1}\, |}
\newcommand{\ket}[1]{|\,{#1}\,\rangle}

\newcommand{\ssection}[1]{{\noi  \it #1:}}
\newcommand{\pdiff}[2]{\frac{\partial #1}{\partial #2}}

\newcommand{\comm}[2]{\big[#1,#2\big]}

\newcommand{\sub}[2]{{#1}_{ \mbox{\scriptsize #2}}}

\newcommand{\bv}[1]{\mathbf{ #1 }}

\def\noi{\noindent}
\def\beq{\begin{equation}}
\def\eeq{\end{equation}}

\def\CR{\nonumber\\[0.15cm]}
\newcommand{\id}{\mathds{1}}
\newcommand{\rref}[1]{Ref.~\cite{#1}}
\newcommand{\frefp}[2]{Fig.~\ref{#1}~(#2)}
\newcommand{\aref}[1]{\ref{#1}}
\newcommand{\bref}[1]{(\ref{#1})}
\usepackage{ulem}  %unter- (\uline{important}) und durchstreichen (\sout{wrong})
\begin{document}

%%%%%%%%%%%%%%%%%%%%%%%%%%%%%%%%%%%%%%%%%%%%%%%%%%%%%%%%%%%%%%%%%

\title{Excitation transport through Rydberg dressing}
\author{S.~W\"uster, C.~Ates, A.~Eisfeld and J.~M.~Rost}
\address{Max Planck Institute for the Physics of Complex Systems, N\"othnitzer Strasse 38, 01187 Dresden, Germany}
\ead{sew654@pks.mpg.de}

%%%%%%%%%%%%%%%%%%%%%%%%%%%%%%%%%%%%%%%%%%%%%%%%%%%%%%%%%%%%%%%%%

%\date{\today}% It is always \today, today,
             %  but any date may be explicitly specified

\begin{abstract}

We show how to create long range interactions between alkali-atoms in different hyper-fine ground states, with the goal of coherent quantum transport. The scheme uses off resonant dressing with atomic Rydberg states. We demonstrate coherent migration of electronic excitation through dressed dipole-dipole interaction by full solutions of models with four essential states per atom and give the structure of the spectrum of dressed states for a dimer. In addition we present an effective (perturbative) Hamiltonian for the ground-state manifold and show that it correctly describes the full multi-state dynamics. We discuss excitation transport in detail for a chain of five atoms. In the presented scheme, the actual population in the Rydberg state is kept small. Dressing offers large advantages over the direct use of Rydberg levels: It reduces ionisation probabilities and provides an additional tuning parameter for life-times and interaction-strengths.
\end{abstract}

\pacs{
32.80.Ee,   % Rydberg States						%| 
82.20.Rp,   % State to state energy transfer 			%| -> these three are as in the cradle paper
34.20.Cf,    % Interatomic potentials and forces	       %|
32.80.Qk 	  %Coherent control of atomic interactions with photons 
}			     
%\maketitle
%
%%%%%%%%%%%%%%%%%%%%%%%%%%%%%%%%%%%%%%%%%%%%%%%%%%%%%%%%%%%%%%%%%
%\section{Introduction}
\section[Introduction]{Introduction\label{introduction}}

Due to their remarkable properties \cite{book:gallagher}, Rydberg atoms emerge as a versatile tool in ultra-cold atomic physics. They can be used for diverse topics, such as quantum information \cite{lukin:quantuminfo,Saffman-M&oslash;lmer-Quantuminformationwith-2010}, atomic aggregates \cite{muelken:exciton:survival}, the study of conical intersections \cite{us:CI}, digital quantum simulations \cite{Weimer-Buchler-Rydbergquantumsimulator-2010}, electro-magnetically induced transparency \cite{adams:cooperative}, coherent population trapping \cite{schempp:poptrap}, giant Kerr non-linearities \cite{adams:giantkerr} and ultra-long range molecules \cite{Greene:LongRangeMols,liu:ultra_long_range_2009, pfau:rydberg_trimers}, since their long range interactions can be tuned over many orders of magnitude. However excited atoms in Rydberg orbitals are more vulnerable to ionisation \cite{Beterov-Bezuglov-IonizationofRydberg-2009} or spontaneous decay and more difficult to trap \cite{Hezel-Schmelcher-ControllingultracoldRydberg-2006,Hezel-Schmelcher-UltracoldRydbergatoms-2007}.

Recently, several groups proposed to combine the advantages of both, ground-state and Rydberg atoms, through ``Rydberg-dressing'' schemes \cite{nils:supersolids,pupillo:strongcorr:dressed,mayle:rydbergtrap:gsprobe,mayle:rydbergtrap:dressing_in_trap,Johnson-Rolston-InteractionsbetweenRydberg-dressed-2010}. The core idea is to use off-resonant laser coupling between ground- and Rydberg states to create eigenstates of the laser-coupled system in which a small Rydberg component is admixed to the ground-state. These resulting \emph{dressed} states inherit some of the extreme properties of the Rydberg states, while preserving favourable properties of the ground state. Moreover, the additional degrees of freedom provided by laser Rabi-frequency and detuning increase flexibility and promise potential shaping as well as dynamical control.

Dressing has been extensively studied for van der Waals (vdW) interactions, generated when virtual transitions between levels of a two-atom system yield a distance dependent energy shift. Several articles have proposed Rydberg dressing techniques to induce this kind of interactions for ground-state atoms \cite{nils:supersolids,pupillo:strongcorr:dressed,mayle:rydbergtrap:gsprobe,mayle:rydbergtrap:dressing_in_trap,Johnson-Rolston-InteractionsbetweenRydberg-dressed-2010,Honer-Buechler-CollectiveMany-BodyInteraction-2010}. In this context it was shown that dressing ground-state atoms with Rydberg levels allows the large Rydberg vdW interaction energies to be reduced until comparable to typical energy scales of cold-atom traps, while simultaneously increasing ensemble life-times to those required for thermal equilibration effects \cite{nils:supersolids,pupillo:strongcorr:dressed}. These are helpful features e.g., for realising a supersolid phase in Rydberg-dressed Bose-Einstein condensates \cite{nils:supersolids,Cinti:Pupillo:SupersolidDropletCrystal:2010}.

Here, we demonstrate that the ideas behind dressing Rydberg vdW interactions can also be applied to resonant dipole-dipole interactions of neutral atoms. They occur when transitions between different states of the two-atom system are energetically resonant. In this case, the existence of an interaction potential is linked to electronic state transfer between atoms \cite{noordam:interactions}. Dressing dipole-dipole interactions requires the use of two laser couplings and four electronic states (two ground- and two Rydberg-states), in contrast to one coupling and two states that suffice for vdW dressing. Hence, the scheme is more involved but also more flexible, offering a larger number of dressed states. It ultimately allows coherent quantum state transport between atoms in different long-lived ground-states over distances of $5$-$15$ $\mu$m and for durations of many milli-seconds.

Similar ideas have previously been applied to trapped Rydberg \emph{ions} \cite{mueller:iondressing}. However, we are motivated by the possibility to generate potentials in an atomic many-body system that induce different kinds of (coupled) electronic and atomic \emph{motion} dependent on the overall quantum state of the system. Useful applications of such potentials can be found in \cite{wuester:cradle,us:CI}. In many systems this goal is easier to achieve with neutral atoms than with ions, since the motion of the latter is typically dominated by their Coulomb repulsion or the indispensable trap. Furthermore the mapping of state-transfer interactions to normally non-interacting ground states by the laser dressing can be presented more clearly in the atomic case, where complications by trapping, mixing of internal and external dynamics and additional laser induced interactions can be ignored at least for free atoms. Finally, see \rref{wall:dressing} for dressing techniques targeting the control of atomic motion without linked excitation transport.

Our results provide an additional handle on the time-scale of excitation transport and life-times, enable time-dependent control over hopping-strengths and can be used to vary the order of magnitude of dipole-dipole forces. Within some constraints \cite{mayle:rydbergtrap:dressing_in_trap}, the dressing also enables the use of well established trapping methods. As recently shown, dressed dipole-dipole interactions are an important tool for the realisation of atomic ring trimers \cite{us:CI}. In particular, the confinement of long-range interacting atoms on a ring, as required in \rref{us:CI}, is greatly facilitated by the techniques presented here. Atomic ring trimers allow the detailed study of wave-function dynamics near and across conical-intersections. Dressed dipole-dipole interactions may also prove useful for the study of exciton dynamics in Rydberg chains \cite{muelken:exciton:survival,cenap:motion,wuester:cradle,moebius:cradle}. 

To demonstrate dressed dipole-dipole interactions, we employ numerical simulations of a model with four essential states per atom. In this case there are several dressed state manifolds, distinguished by the number of excited Rydberg atoms which they contain in the limit of vanishing dressing. We apply Van-Vleck perturbation theory to obtain analytical expressions for the induced effective interaction and to determine the parameter range within which the scheme can function. 

The paper is organised as follows: In \sref{system} we present the atomic states considered, describe the resulting Hamiltonian and outline the basic principles of the dressing scheme. In \sref{dimer} we consider an atomic dimer in detail, providing analytical expressions for the effective dressed interactions, diagonalising the full Hamiltonian and discussing different available dressing manifolds. In \sref{applications} we present two exemplary applications of our results, dressed ring trimers and exciton migration. We conclude in \sref{conclusion} and discuss technical details of our calculations and the underlying atomic physics in the subsequent appendices.

%%%%%%%%%%%%%%%%%%%%%%%%%%%%%%%%%%%
%%%%%%%%%%%%%%%%%%%%%%%%%%%%%%%%%%%
\section{Dressed Rydberg aggregates} 
%%%%%%%%%%%%%%%%%%%%%%%%%%%%%%%%%%%
%%%%%%%%%%%%%%%%%%%%%%%%%%%%%%%%%%%
\label{system}

%%%%%%%%%%%%%%%%%%%%%%%%%%%%%%%%%%%
\subsection{Model}
%%%%%%%%%%%%%%%%%%%%%%%%%%%%%%%%%%%
\label{model}
\begin{figure}[htb]
\psfig{file=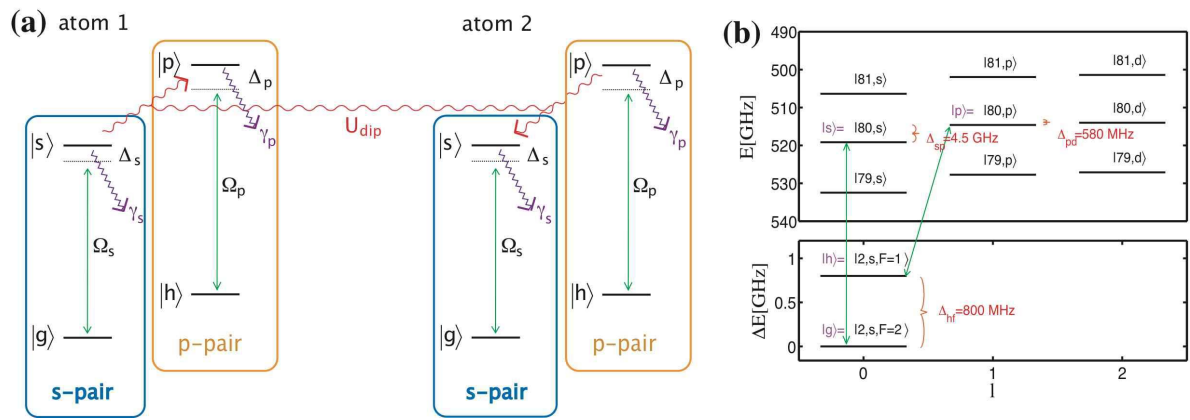,width=\columnwidth}
\caption{\label{fig:sketch}(a) Schematic level diagram for laser dressing of a pair of ground state atoms with Rydberg states. The ground states $\ket{g}$, $\ket{h}$ do not participate in inter-atomic interactions nor spontaneously decay on time-scales of interest. States $\ket{s}$, $\ket{p}$ are highly excited Rydberg states, participating in binary long range interactions as explained in the text. Ground and excited states are coupled in a far detuned fashion as indicated in the diagram. The symbol $\gamma$
indicates the relevance of spontaneous decay, which we discuss in \sref{decay}. (b) Implementation of the scheme sketched in (a) for $^7$Li, using the indicated states to realise $\ket{g}$, $\ket{h}$, $\ket{s}$, $\ket{p}$. Also shown are the states energetically closest to $\ket{s}$, $\ket{p}$. For the hyperfine-split ground state, $F$ denotes the total atomic angular momentum (nuclear, orbital and spin). See \aref{full:levels} for further details regarding the indicated transitions.}
\end{figure}
We consider a collection of $N$ alkali atoms, where the $n$'th atom is located at the position $\bv{R}_{n}$. The distances between the atoms, $R_{kl}=|\bv{R}_{k} - \bv{R}_{l}|$, shall be so large that interaction between two atoms can be neglected unless \emph{both} of them are in a Rydberg state. In this case they experience strong dipole-dipole interactions. As discussed in \cite{muelken:exciton:survival,noordam:interactions,cenap:motion,wuester:cradle,moebius:cradle} the dipole-dipole interaction leads to exciton dynamics similar to that of molecular dye aggregates, molecular crystals or light harvesting systems \cite{AmVaGr00__}. 

For practical reasons we restrict ourselves to situations where the dynamics can be adequately described using only two ``ground'' states and two Rydberg states per atom. With the term ``ground-states'' we refer e.g., to two different hyperfine levels of the actual alkali-atom ground state.  We shall denote the ground states by $\ket{g}$ and $\ket{h}$ and the Rydberg states by $\ket{s}$ and $\ket{p}$. The latter correspond to the $l=0$ and $l=1$ angular momentum components of a Rydberg level with large principal quantum number~$\nu \gtrsim 30$. Schematically, the resonant dipole-dipole interaction $\hat{U}(\bv{R})$ leads to a coupling between states $\ket{sp}$ and $\ket{ps}$, whose matrix elements in the Hamiltonian take the form $\sub{U}{dip}(R)\ket{sp}\bra{ps}$.
The transition strength is $\sub{U}{dip}(R)=-\mu^2/R^3,$ where $R$ is the inter-atomic distance and $\mu$ quantifies the transition dipole moment between the states $\ket{s}$ and $\ket{p}$. We assume that all atoms are prepared in the $m_{l}=0$ azimuthal quantum states and never acquire $m_{l}\neq0$. With atoms constrained in a two-dimensional (2D) plane, orthogonal to the quantisation axis, this ensures that there is no angular dependence of $\hat{U}(\bv{R})$~\cite{noordam:interactions,moebius:cradle}. Such a geometry covers both scenarios described later in this article. The interaction between atoms $k$ and $l$ reads 
\begin{equation}
U_{kl}=\sub{U}{dip} (R_{kl})=-\frac{\mu^2}{|\bv{R}_{k} - \bv{R}_{l}|^3}.
\end{equation}
To map the strong interactions $U_{kl}$ to the ground states, the atoms are irradiated with far-detuned dressing laser-fields that couple the ground-states and our two selected Rydberg states coherently. The relevant level diagram is sketched in \frefp{fig:sketch}{a}. In practice such a coupling is commonly achieved by two (or multi)-photon transitions \cite{book:demtroeder}, so that the Rabi-frequencies $\Omega_{s,p}$ and detunings $\Delta_{s,p}$ in the diagram have to be regarded as effective quantities. From the laser transition parameters, we assemble the effective dimensionless dressing parameters 
\begin{eqnarray}
\label{alphasp}
\alpha_{s,p}=\frac{\Omega_{s,p}}{2\Delta_{s,p}}.
\end{eqnarray}
They are a measure of how ``far-detuned'' the laser coupling is, and will emerge as crucial quantities controlling the dressing, as it is the case for dressed vdW interactions \cite{nils:supersolids,pupillo:strongcorr:dressed}.
It is important that the state $\ket{g}$ is directly coupled only to $\ket{s}$ and $\ket{h}$ only to $\ket{p}$. In order for this simple picture with four relevant states per atom to be valid, the detunings of both laser couplings have to be chosen such that all other transitions, coupling to further Rydberg states or ground state levels, are so far detuned that they can be safely neglected\footnote{If both couplings are realised by two-photon transitions, these considerations should include the virtual middle level. We require couplings $\ket{g}\leftrightarrow\ket{m_{s}} \leftrightarrow \ket{s}$ and $\ket{h}\leftrightarrow\ket{m_{p}} \leftrightarrow \ket{p}$ with uniquely assigned states $\ket{m_{s/p}}$. See \aref{full:levels} for more details.
}.
 We show some realistic level diagrams for $^7$Li in \frefp{fig:sketch}{b} to demonstrate how this constraint can be met in practice. Throughout the article, we will refer to the  states $\ket{g}$, $\ket{s}$ as the ``s-pair'' and the states $\ket{h}$, $\ket{p}$  as the ``p-pair''. 

Using the four states introduced above as a basis for the single atom, an $N$-body basis state $\ket{\bv{k}}$ is written as 
\begin{eqnarray}
\label{nbodybasis}
\ket{\bv{k}}\equiv \ket{k_1\dots k_N}\equiv\ket{{k_{1}}} \otimes \dots \otimes\ket{{k_{N}}},
\end{eqnarray}
 where $k_{j}\in \{{g},{h},{s},{p}\}$ describes the electronic state of the atom $j$. For example we write $\ket{ghs}$ when the first atom is in state $\ket{g}$, the second in $\ket{h}$ and the third in $\ket{s}$. After defining operators $\hat{\sigma}^{(n)}_{kk'}=\ket{k_{n}}\bra{k'_{n}}$ with $k$, $k'$ $\in \{g,h,s,p \}$ where $n$ is the atom-index, the many-body Hamiltonian can be written as 
\begin{eqnarray}
\label{Hamil_dressed}
\hat{H}=\hat{H}_{0} + \hat{V},
\end{eqnarray}
with
\begin{eqnarray}
\label{Hunpert}
\hat{H}_{0}=-\Delta_{s}\sum_{n}\hat{\sigma}^{(n)}_{ss} -\Delta_{p}\sum_{n}\hat{\sigma}^{(n)}_{pp}  +\sum_{nl}U_{nl}\hat{\sigma}^{(n)}_{sp}\hat{\sigma}^{(l)}_{ps},
\end{eqnarray}
and
\begin{eqnarray}
\label{Hpert}
\hat{V}=\sum_{n}\left(\frac{\Omega_{s}}{2}\hat{\sigma}^{(n)}_{gs}  +  \frac{\Omega_{s}^{*}}{2}\hat{\sigma}^{(n)}_{sg}\right) + \sum_{n}\left(\frac{\Omega_{p}}{2}\hat{\sigma}^{(n)}_{hp}  +  \frac{\Omega^{*}_{p}}{2}\hat{\sigma}^{(n)}_{ph} \right).
\end{eqnarray}
We split $\hat{H}$ into two parts, $\hat{H}_{0}$ and $\hat{V}$, to facilitate our later application of perturbation theory. 
The atom-light coupling in \eref{Hamil_dressed} has been treated in the dipole- and rotating wave approximations, as explained in \aref{rotwave}. 
The \emph{exciton number} operator
\begin{eqnarray}
\label{Nexcitons}
\hat{N}_{e}=\sum_{n}\left(\hat{\sigma}^{(n)}_{pp}  +  \hat{\sigma}^{(n)}_{hh}   \right),
\end{eqnarray}
measures the number of atoms in the p-pair of states. It is easy to see that $\hat{N}_{e}$ commutes with $\hat{H}$ and hence the exciton number is conserved, see also \aref{multi_exc}.

Of course equation \eref{Hamil_dressed} is a simplified description of a complicated multi-state system. For a more complete picture, one could include additional Rydberg levels adjacent to $\ket{s}$, $\ket{p}$, as shown in \frefp{fig:sketch}{b}, van der Waals interactions of all four states, spontaneous decay and loss in the excited states. The latter arises as atoms in Rydberg states are lost due to auto-ionization or blackbody radiation \cite{beterov:BBR,theodosiou:lifetime}. 
Importantly, parameter regimes can be found where these corrections are minor and \eref{Hamil_dressed} suffices, for example for the states and parameters shown in \frefp{fig:sketch}{b} and described in \sref{dressingintro}. 

%%%%%%%%%%%%%%%%%%%%%%%%%%%%%%%%%%%
\subsection{Basic principles and exemplary parameters}
%%%%%%%%%%%%%%%%%%%%%%%%%%%%%%%%%%%
\label{dressingintro}

Consider an atomic sample such as a chain, prepared in states $\ket{\pi_{n}}$, in which the n'th atom is in the upper ground state $\ket{h}$ and all others in the absolute ground-state $\ket{g}$ 
\begin{equation}
\ket{\pi_{n}}=\ket{ggg...h...ggg}.
\end{equation}
 Without dressing ($\Omega_{s}=\Omega_{p}=0$) this is an eigenstate of the Hamiltonian \bref{Hamil_dressed}. If we ignore dipole-dipole interactions for the moment, an adiabatic ramp-up of the dressing strengths $\Omega_{s,p}$ will change this eigenstate into one with some admixture of Rydberg population. Allowing then for dipole-dipole interactions finally causes excitation transport via processes of the schematic form: $\ket{gh}\rightarrow  \ket{sp}\rightsquigarrow  \ket{ps} \rightarrow  \ket{hg}$. Here $\rightarrow$ symbolizes a transition of both atoms between the ground state manifold and the Rydberg state manifold due to the dressing lasers. Similarly $\rightsquigarrow$ stands for a transition occurring due to resonant dipole-dipole interactions between Rydberg states. In \sref{dimer} we will confirm this expectation with numerical simulations and later justify and quantify it using perturbation theory. If the lasers are sufficiently off-resonant, the admixed fraction of excited state should remain small throughout. For this reason we can describe the process that we have just sketched as an effective interaction in the space spanned by the $\ket{\pi_{n}}$. We will refer to it as the \emph{ground-state manifold}. For later reference we also define spaces spanned by  
\begin{eqnarray}
\ket{\pi_{n}}_{A}=\ket{ggg...p...ggg},
&\:\:\:\:\:\:\: \ket{\pi_{n}}_{B}=\ket{hhh...s...hhh},
\end{eqnarray}
which we call \emph{gp (hs) single excitation manifold}.

Throughout this paper we focus on a specific realisation of our scheme, sketched in \frefp{fig:sketch}{b}. We consider $^7$Li atoms, whose $^2 S_{1/2}$ ground-state has a hyperfine-splitting of $\Delta_{hf}=800$ MHz \cite{beckmann:LiDeltaHF}. For the Rydberg state $\ket{s}$ we pick a principal quantum number $\nu=80$ and angular momentum $l=0$ state. For $\ket{p}$ we choose $\nu=80$ and $l=1$. The energy difference between the two Rydberg states is $4.5$ GHz. We will assume detunings $\Delta_{s}=\Delta_{p}=\Delta=50$ MHz and dressing parameters $\alpha_{s}= \alpha_{p}= \alpha=0.02$ unless indicated otherwise. 
We will occasionally refer to $\alpha$ (without index), which implies $\alpha_{s}= \alpha_{p}= \alpha$.
For the above values of $\alpha$ and $\Delta$, our Rabi-frequencies are $\Omega_s=\Omega_p=4$ MHz. 

For the most part, we illustrate the dressing scheme through an atomic dimer with $R=6.5$ $\mu$m inter-particle separation. For this distance, the strength of the bare dipole-dipole interaction is $U=92.5$~MHz, just less than the doubly excited state detuning $\Delta_{s} + \Delta_{p}= 100$ MHz. The reason for this choice will become clear later. 

We stress that we have chosen these parameters to be specific, while our findings are more general. The constraints under which the parameters can be varied without invalidating the dressing scheme have been discussed in \sref{model}. All following results are expected to hold for any alkali species if these constraints are met. 
 
%%%%%%%%%%%%%%%%%%%%%%%%%%%%%%%%%%%
%%%%%%%%%%%%%%%%%%%%%%%%%%%%%%%%%%%
\section{Dressed dimer}
%%%%%%%%%%%%%%%%%%%%%%%%%%%%%%%%%%%
%%%%%%%%%%%%%%%%%%%%%%%%%%%%%%%%%%%
\label{dimer}

For the principal demonstration of excitation transport through Rydberg dressing, we first treat the simplest possible case, the atomic dimer. 

%%%%%%%%%%%%%%%%%%%%%%%%%%%%%%%%%%%
\subsection{Dressing dynamics in the ground state manifold}
%%%%%%%%%%%%%%%%%%%%%%%%%%%%%%%%%%%
\label{dressingdemo}

\begin{figure}[htb]
\epsfig{file=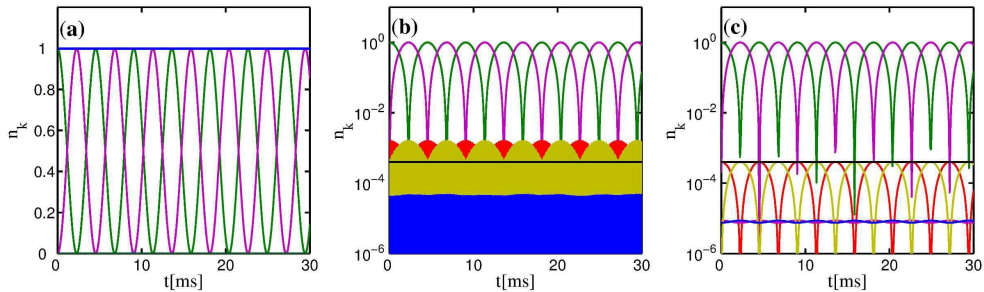,width=\columnwidth}
\caption{\label{N2_all} Transport dynamics of a Rydberg-dressed dimer, (a) Populations $n_{k}=|c_{k}|^2$ of the two-atom states $\ket{gh}$ (violet) and $\ket{hg}$ (green) on a linear scale. The blue line is the total population in the ground state manifold. (b,c) On a logarithmic scale the differences
of the two ramp-ups become visible: (b) sudden addition of the laser coupling: The singly excited states (red, yellow) are roughly suppressed by $\alpha^2=4\times 10^{-4}$, indicated by the solid black line.
Doubly excited states (blue) are even further suppressed. (c) adiabatic ramp-up of the laser coupling: singly excited state populations are precisely suppressed by $\alpha^2$. The expected spontaneous lifetime of this system is $\sub{\tau}{eff}=(\alpha^2(\tau_{s}^{-1} +  \tau_{p}^{-1}))^{-1}=290$ms as will be described in \sref{decay}.
}
\end{figure}
In this section, we show dynamics from the initial state $\ket{\pi_{1}}=\ket{hg}$. We consider two different ramps of the dressing couplings $\Omega_{s,p}$ in the Hamiltonian: (i) a sudden jump from zero to their final value and, (ii) an adiabatic ramp over a small finite time $\sub{T}{ramp}=0.1$ ms, effectively changing the initial state to $\ket{\tilde{\pi}_{1}}\equiv (\ket{hg} + {\cal O}(\alpha_{s,p})[\ket{pg} + \ket{hs}])/{\cal N}$, where ${\cal N}$ is a normalisation factor (see also \sref{dimer_vanVleck}).
\footnote{
To obtain dressed excitation transport from a state $\ket{\tilde{\pi}_{1}}$, the ramp ought to be slow on the time-scale of the laser-coupling but fast on that of effective dipole-dipole transport. Thus $\sub{T}{dress}<\sub{T}{ramp}<\sub{T}{ex}$ with $\sub{T}{dress}=2\pi/\Omega_{s,p}=0.25$ $\mu$s and $\sub{T}{ex}\approx4$ ms (The value is read off \fref{N2_all} and defined as the time it takes for the excitation to oscillate from the first atom to the second and back.).
}.

We expand the time dependent quantum state as $\ket{\Psi(t)}=\sum_{\bv{k}} c_{\bv{k}} \ket{\bv{k}}$, where $\ket{\bv{k}}$ are the $N$-body basis vectors defined in \eref{nbodybasis}. In this basis, we express the Hamiltonian in matrix form and solve the Schr{\"o}dinger equation (SE) $i \partial_{t} \ket{\Psi(t)} = \hat{H}\ket{\Psi(t)}$ by matrix exponentiation. Using the four-state model introduced in \sref{model} we employ the full Hamiltonian \eref{Hamil_dressed} without further approximations.

For two atoms placed a distance $R=6.5$ $\mu$m apart, the resulting population-oscillations between the states $\ket{gh}$ and $\ket{hg}$ can be clearly seen in \fref{N2_all}. On the logarithmic scale of panel (b), we see that also states containing one Rydberg excitation or more are populated, but their occupation is roughly suppressed by $\alpha^2=4\times 10^{-4}$, indicated by the black solid line. The suppression of excited state population by a factor $\alpha^2$ is well known from vdW dressing techniques \cite{nils:supersolids,pupillo:strongcorr:dressed} and a basic feature of off-resonant Rabi-coupling. As can be seen in \fref{N2_all}, there will be many ground state population oscillations up to the expected spontaneous life-time of the dressed $2$-atom state, which we estimate to about $0.29$s as described in \sref{decay}. Due to the dressing, the time-scale on which the interaction transports the excitation has been massively increased compared to the ``bare'' population oscillation period between Rydberg states $\ket{sp}\leftrightarrow \ket{ps}$, which would be $\sub{T}{ex}=5$ns. 

The comparison of panels (b) and (c) of \fref{N2_all} shows that the use of the more refined, adiabatically created, initial state has led to a more regular evolution of the excited state populations, whose oscillations now precisely follow those of the two ground states. The population of states with a single Rydberg excitation never exceeds $\alpha^2$. Doubly excited states are further suppressed. However, this careful initial state creation is not actually required in order to observe dressed population oscillations, as we have just seen.

%%%%%%%%%%%%%%%%%%%%%%%%%%%%%%%%%%%
\subsection{Effective Hamiltonian}
%%%%%%%%%%%%%%%%%%%%%%%%%%%%%%%%%%%
\label{dimer_vanVleck}

The results of the preceding section already unambiguously demonstrate excitation transport through full numerical solutions of the problem.
To further understand the population-oscillations more intuitively, we now consider the ground-state manifold spanned by $\ket{\pi_{1}}$, $\ket{\pi_{2}}$ as the \emph{system} of interest and its Hilbert-space complement as \emph{environment}. The coupling between the two is treated as a perturbation. We use van Vleck perturbation theory \cite{vanVleck:PT}, as outlined by Shavitt {\it et al.} \cite{shavitt:vanVleck} to derive an effective Hamiltonian in terms of the ``system'' only. This scheme conveniently takes care of degeneracies and generates an effective Hamiltonian well suited to describe excitation transport. A full analytical diagonalisation of \eref{Hamil_dressed} is impractical even for $N=2$. In contrast the perturbative results for $N=2$ are intuitive and generalise straightforwardly to cases with more atoms. They are also valid over large ranges of parameter space as we shall show below. 

The aim of van Vleck perturbation theory is to find a basis that block-diagonalises the Hamiltonian \bref{Hamil_dressed} to a given order in the perturbation $\hat{V}$. For $\hat{V}=0$ this is achieved by the basis $\ket{\bv{k}}$ introduced in \sref{system}, since $\hat{H}_{0}$ is already fully diagonal. The relative importance of $\hat{V}$ and $\hat{H}_{0}$ is governed by the dressing parameters $\alpha_{s,p}$. As the $\alpha_{s,p}$ increase, the basis that block-diagonalises $\hat{H}$ becomes more complicated. Perturbation theory will thus be valid as long as the dressing parameters $\alpha_{s,p}$ are small.

We will use the notation 
\begin{eqnarray}
\ket{\bv{k}} \rightarrow \ket{\tilde{\bv{k}}}=\ket{\bv{k}} +  \sum_{\bv{m}}b_{\bv{m}}\ket{\bv{m}}, 
\end{eqnarray}
with $b_{\bv{m}} \sim {\cal O}(\alpha_{s,p})$, i.e.~$\ket{\tilde{\bv{k}}}$ is a state whose leading component is $\ket{\bv{k}}$ as long as $\alpha$ remains small and $U$ is small (i.e.~R is large). In this section we shall only be interested in the block of the effective Hamiltonian that governs the dynamics of the effective ground-state manifold $\ket{\tilde{\pi}_1}=\ket{\tilde{hg}}$ and $\ket{\tilde{\pi}_2}=\ket{\tilde{gh}}$. The details of the calculation are given in \aref{vanvleck}, here we merely present the results. In the basis $\ket{\tilde{\pi}_1}$, $\ket{\tilde{\pi}_2}$ one obtains
\begin{eqnarray}
\sub{H}{eff}=(E_{2} + E_{4})\id + E(R),
\label{H_eff}
\end{eqnarray}
where $\id$ is the $2\times2$ identity matrix. We have energy shifts $(E_{2} + E_{4})$ and an explicitly distance dependent part $E(R)$. The light shifts $E_{2}$ ($E_{4}$),
corresponding to second (fourth) order in $\hat{V}$, are given by
\begin{subequations}
\label{Eshifts}
\begin{eqnarray}
E_{2} =  \alpha_{s}^2 \Delta_{s}+  \alpha_{p}^2\Delta_{p}, 
\end{eqnarray}
\begin{eqnarray}
E_{4} = -(\alpha_{s}^4 \Delta_{s}+  \alpha_{p}^4\Delta_{p} +   \alpha_{s}^2 \alpha_{p}^2(\Delta_{s} + \Delta_{p}) ), 
\end{eqnarray}
\end{subequations}
where we have used the dressing parameters defined in \eref{alphasp}. These light shifts could allow optical trapping using the same fields that provide the dressing coupling. The position dependent part of the Hamiltonian, $E(R)$, can be written as
\begin{eqnarray}
E(R)=
\left(
\begin{array}{ccc}
W(R) & \tilde{U}_{12}(R)   \\
\tilde{U}_{21}(R) & W(R)  \\
\end{array}
\right),
\label{ER}
\end{eqnarray}
with
\begin{subequations}
\label{potentials}
\begin{eqnarray}
W(R) =  \alpha_{s}^2 \alpha_{p}^2 \frac{1}{1 -  \sub{ \bar{U} }{12}^2(R) }(\Delta_{s} + \Delta_{p}), 
\label{diagpotential}
\end{eqnarray}
\begin{eqnarray}
\tilde{U}_{12}(R)  =  \alpha_{s}^2 \alpha_{p}^2 \frac{ \sub{U}{12}(R)  }{1 - \sub{\bar{U}}{12}^2(R) },
\end{eqnarray}
\end{subequations}
where we have defined $\sub{ \bar{U} }{ij} =\sub{U}{ij} /(\Delta_{s} +\Delta_{p})$.  As expected the states $\ket{\tilde{\pi}_{1}}$, $\ket{\tilde{\pi}_{2}}$ acquire small excited state populations of the form:
\begin{eqnarray}
\ket{\tilde{\pi}_{1}}=(\ket{hg} + \alpha_{p} \ket{pg} + \alpha_{s} \ket{hs} + {\cal O}(\alpha_{s,p}^2) )/{\cal N},
\label{effective_states}
\end{eqnarray}
independent of $R$. Here ${\cal N}=\sqrt{1 +  \alpha_{p}^2 + \alpha_{s}^2}$ is a normalisation factor. 

Let us briefly mention some special cases: For large inter-atomic separations, such that $\sub{U}{ij} \ll (\Delta_{s} +\Delta_{p})$, we can expand the expressions \bref{potentials} and obtain $W(R) \sim W_{\infty}= -\alpha_{s}^2 \alpha_{p}^2 (\Delta_{s} + \Delta_{p})$ and $U(R) \sim
\alpha_{s}^2 \alpha_{p}^2  \sub{U}{12}$. The shift $W_{\infty}$, which is independent of $R$, can then be merged with $E_{4}$. For shorter distances, the factors $1/(1 - \sub{\bar{U}}{12}^2)$ become relevant and diverge at $\sub{\bar{U}}{12}=1$, or equivalently at $\sub{U}{ij}(R) = (\Delta_{s} +\Delta_{p})$. We will see in the next section that this divergence can be traced to avoided crossings between the perturbed energy eigenvalues. We show in \sref{checks} that \bref{potentials} fails only in a fairly narrow region around this avoided crossing. 

An interesting special case is $\Delta_s=-\Delta_p$. Here, the equations predict that the dressing effect on the ground-state manifold vanishes, which is confirmed by full simulations. We will however show in \sref{singleexc} that this case has special appeal for dressing within the single excited state manifold.

We also derive the contribution to $\sub{H}{eff}$ of $6$'th order in $\hat{V}$ in \aref{sixth_order}. We will make use of it in \sref{checks}, to illustrate the rate of improvement that can be achieved through higher-order terms of perturbation theory.

%%%%%%%%%%%%%%%%%%%%%%%%%%%%%%%%%%%
\subsection{Potential surfaces}
%%%%%%%%%%%%%%%%%%%%%%%%%%%%%%%%%%%
\label{surfaces}

To get an overview of the consequences of dressing radiation beyond the ground state manifold, we now consider the full energy spectrum of the dimer as a function of inter-atomic separation. As we outline in \aref{multi_exc}, the dimer Hamiltonian can be brought into block-diagonal form, with three blocks describing exciton numbers of $N_{e}=0,1,2$. Of these, only the single exciton block is non-trivial, hence we only show the eigenspectrum of its corresponding sub-matrix $M_{1}$ \bref{Hamiltonian_mainblock} here. Explicitly we solve 
\begin{eqnarray}
M_{1}\ket{\Psi_{k}} = E_{k}\ket{\Psi_{k}}.
\label{eigenvalue_eqn}
\end{eqnarray}
Besides providing information on the possibilities of dressed excitation transport, the energy eigenvalues as a function of separation can also be viewed as adiabatic (Born-Oppenheimer) potential surfaces for motion of the atoms, as long as this motion is sufficiently slow \cite{domke:yarkony:koeppel:CIs}. 

\begin{figure}[htb]
\psfig{file=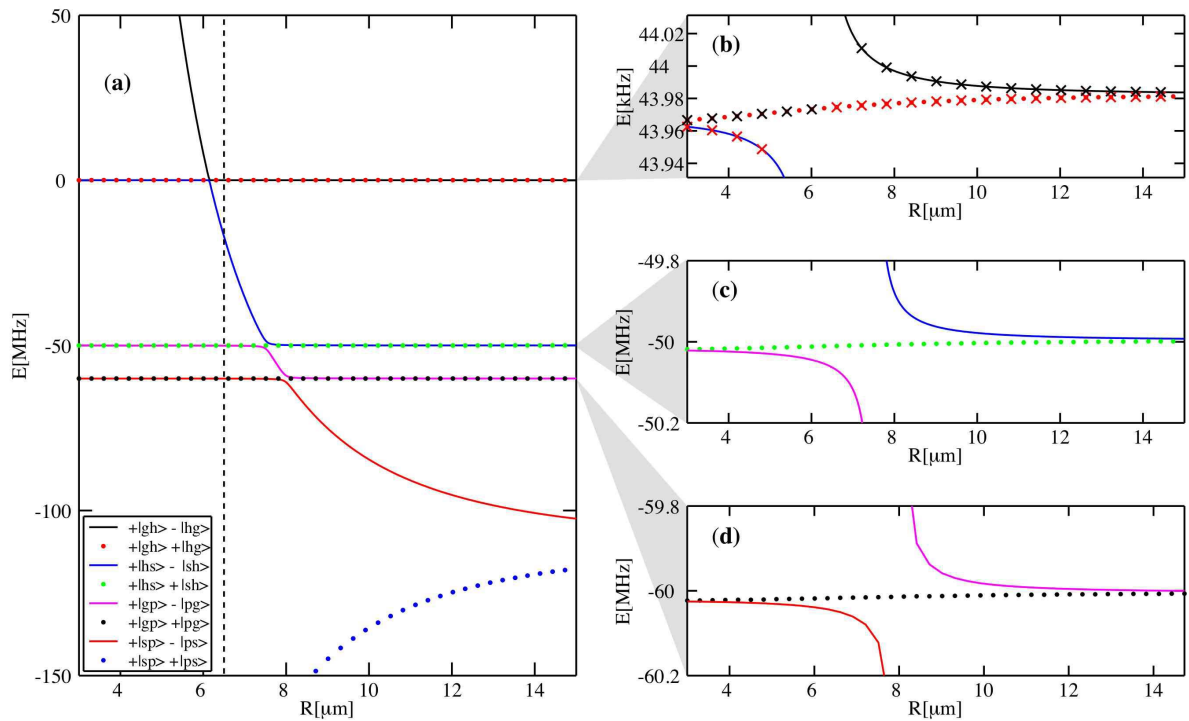,width=\columnwidth}
\caption{\label{spectrum_deltaeq}Energy spectrum of a Rydberg-dressed atomic dimer as a function of atomic separation $R$ for $\Delta_{s}=50$~MHz and $\Delta_{p}=60$~MHz. We plot only energies in the single exciton sub-space, governed by the Hamiltonian \bref{Hamiltonian_mainblock}. All states with odd particle-exchange parity correspond to solid lines, those with even parity to bullets. (a) View on the largest energy-scale set by the detuning. The two states with dominant components $\ket{sp}\pm\ket{ps}$ for large $R$ are most strongly affected by the dipole-dipole interaction. As a consequence, one of them undergoes avoided crossings with other states. Three avoided crossings can clearly be seen. The states are labelled in the legend according to their leading two-body content at large $R$. The state-content character changes each time a state undergoes an avoided crossing, so that for example the solid black curve corresponds to $\ket{gh}-\ket{hg}$ at large $R$, but has become close to $\ket{sp}-\ket{ps}$ at small $R$. The vertical dashed line indicates the separation $R=6.5\mu$m, chosen for dynamical examples throughout this article. (b) Zoom onto ground-state manifold. The crosses indicate the eigenvalues of the effective Hamiltonian \bref{H_eff}. (c-d) Zoom onto singly excited state manifolds, the colour code is as in panel (a). Note the different energy scales of (b) compared to (c) and (d).
}
\end{figure}

The $R$-dependent energy eigenvalues are shown in \fref{spectrum_deltaeq} for $\Delta_{s}=50$~MHz and $\Delta_{p}=60$~MHz, with other parameters as described in \sref{dressingintro}. The slight offset between $\Delta_{s}$ and $\Delta_{p}$ makes it easier to grasp the structure of the resulting spectrum. 
At large distances $R$ and for moderate dressing strength $\alpha$, the eigenstates are essentially superpositions of two basis states, and have odd or even symmetry under exchange of atoms $1$ and $2$. These superpositions are listed in the caption of \frefp{spectrum_deltaeq}{a}, disregarding normalisation. Their energies are then mostly determined by the total detuning of the two superimposed basis states. As the distance $R$ is reduced, the dipole-dipole interaction energy of the states consisting mainly of two Rydberg atoms (blue dotted- and red solid lines) becomes more prominent and eventually leads to avoided crossings between these and the other states of the spectrum. 
On the large energy scale of \frefp{spectrum_deltaeq}{a}, ${\cal O}(\Delta_{s,p})$, the dipole-dipole shift of the $\ket{\tilde{sp}}\pm\ket{\tilde{ps}}$ states is most prominent. However also the other states with only a small doubly excited state component  acquire a space dependent dipole-dipole potential, as can be seen in the close-ups, panels (b)-(d). The energy splitting of the states within the ground-state manifold, panel (b), is of order ${\cal O}(\alpha^4 \Delta_{s,p})$ while that in the singly excited state manifolds, panels (c-d), is of order ${\cal O}(\alpha^2 \Delta_{s,p})$. In panel (b) we additionally display the energy eigenvalues obtained from the effective Hamiltonian \bref{H_eff}. The shape of the potential is well reproduced.

The dressed potentials shown in \fref{spectrum_deltaeq} (b)-(d) have at least two interesting features: (i) Their overall strength scales like $\sim \alpha^2$ or $\sim \alpha^4$, it can thus easily be adjusted by choice of laser parameters and even manipulated time-dependently. (ii) The shape of the potentials can be modified beyond a simple attractive or repulsive form $\sim\pm R^{-3}$, due to the appearance of avoided crossings. For example the attractive branch of potentials in the ground-state manifold, shown in \frefp{spectrum_deltaeq}{b}, approaches a finite value for small separations $R$ as can be seen from its continuation in \frefp{spectrum_deltaeq}{a}. This could avoid acceleration to high velocities during atomic collisions on the attractive potential, maybe reducing the probability of collisional ionisation \cite{amthor:vanderwaals,cenap:motion}. Light induced modifications of potentials that would be strongly attractive in the absence of dressing have already been discussed for the case with a single ground state in \rref{wall:dressing}, and proposed as collisional shield for cold polar molecules in \rref{Gorshkov:molecule:shield}.
However, in our case, once the separation of the atoms is much smaller than the minimal value included in \fref{spectrum_deltaeq}, which is  $R\sim4{\mu}m$, we have to keep in mind that the simplified four-state model breaks down: The dipole-dipole shift of the dressed states then may become so large that they couple strongly to two-atom states that are not included in our model.

\begin{figure}[htb]
\psfig{file=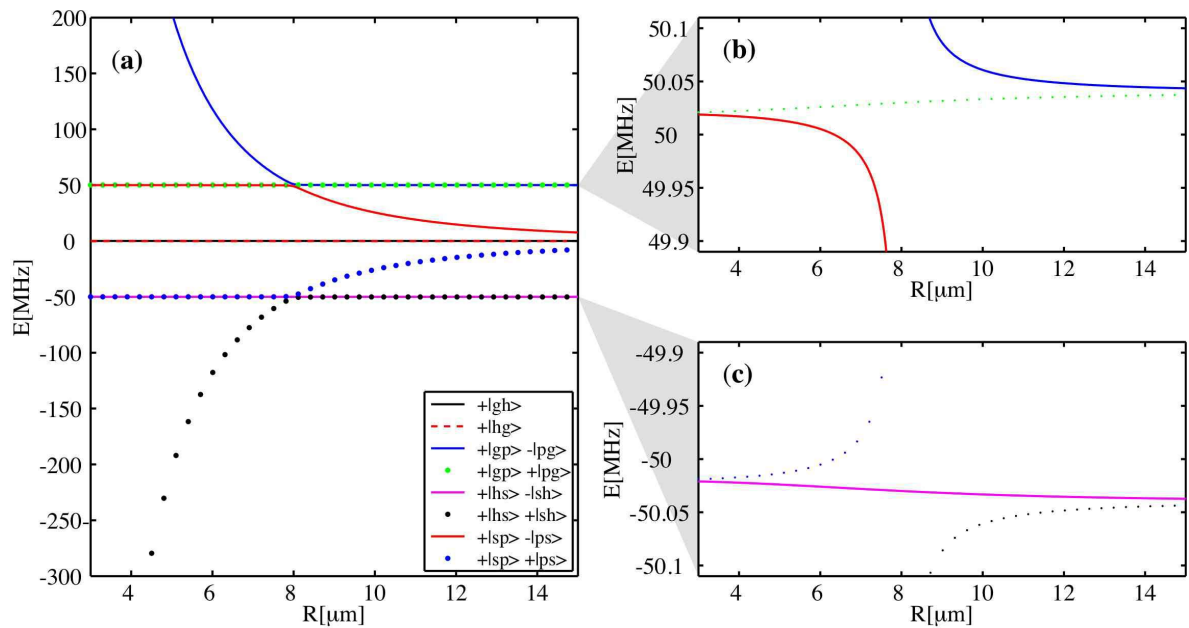,width=\columnwidth}
\caption{\label{spectrum_deltadiff}The same as \fref{spectrum_deltaeq}, but for $\Delta_{s}=-\Delta_{p}=50$MHz. In contrast to \fref{spectrum_deltaeq} the ground-state manifold does not acquire a potential through the dressing. Symmetrical dressed potentials are obtained for the singly-excited state manifolds. The adiabatic potentials for the asymptotic $\ket{\tilde{sp}}\pm\ket{\tilde{ps}}$ states are cut-off at energy $\sim\pm\Delta$ due to avoided crossings.
}
\end{figure}

Let us briefly discuss the case when the detunings have equal magnitudes. As $\Delta_{s}$ approaches $\Delta_{p}$, the asymptotic $\ket{\tilde{gp}}-\ket{\tilde{pg}}$ energy curve (violet line) in \frefp{spectrum_deltaeq}{a} is squeezed between the neighbouring ones until the states become degenerate. If we change the sign of one of the detunings and consider the special case $\Delta_{s}=-\Delta_{p}$, we obtain a particularly symmetric energy spectrum, shown in \fref{spectrum_deltadiff}.  As predicted by \eref{potentials}, the dressing-induced state transfer interaction vanishes between the states $\ket{\tilde{gh}}$ and $\ket{\tilde{hg}}$. However symmetrical induced potentials can now be found in the singly-excited state manifold. The strength of these potentials scales as $\alpha^2$, different from those previously obtained in the ground-state manifold. We will discuss this further in \sref{singleexc}. 

%

%%%%%%%%%%%%%%%%%%%%%%%%%%%%%%%%%%%
\subsection{Performance of effective Hamiltonians}
%%%%%%%%%%%%%%%%%%%%%%%%%%%%%%%%%%%
\label{checks}

\begin{figure}[htb]
\centering
\psfig{file= 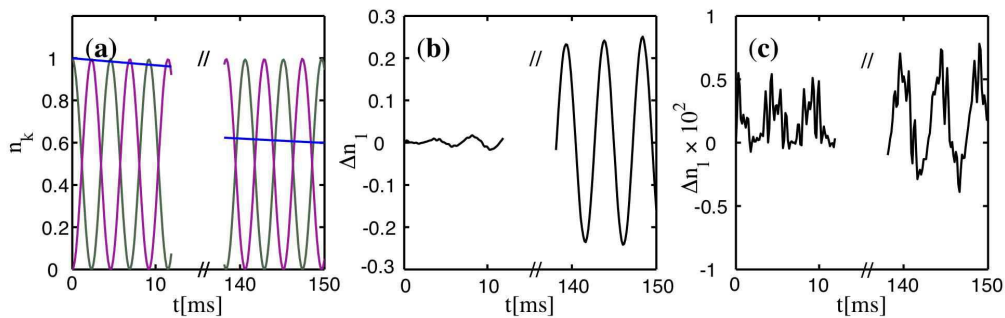,width=\columnwidth}
\caption{\label{comparison_PT_fullH} Comparison between exact dynamics using the Hamiltonian \bref{Hamil_dressed} (a) and the errors resulting from using its perturbative approximations to fourth order (b, see \eref{H_eff}) and sixth order (c, see \eref{ER6}) for the transport dynamics of Rydberg-dressed dimer from a dressed localized state. Panels (b) and (c) show the difference between the respective perturbation orders and the exact solution. Simulations begin from the state $\ket{hg}$ at $t=-\sub{T}{ramp}$. Then, over a duration $\sub{T}{ramp}=100{\mu}$s, we linearly ramp up the Rabi-frequencies $\Omega_{s,p}$ to their final values. This approximately creates the state $\ket{\tilde{hg}}$ at $t=0$.  The colour scheme is as in \fref{N2_all}. The solid blue line in panel (a) shows the function $p(t)=N(t)/N(0)=\exp{(-t/\sub{\tau}{eff})}$, with $\sub{\tau}{eff}=290$ ms for orientation, see discussion in the next section.}
\end{figure}

In this section we revisit dressed dimer dynamics first presented in \sref{dressingdemo}, to apply the results of \sref{dimer_vanVleck}. Figure \ref{comparison_PT_fullH} shows a comparison of the evolution governed by the effective perturbative Hamiltonian \eref{H_eff} with that following from the full Hamiltonian \eref{Hamil_dressed}. Already fourth order perturbation theory describes the effective induced hopping period fairly well, as can be seen in panel (b). When we move to the sixth order expressions, which can be found in \aref{vanvleck}, the agreement becomes even better: The deviation never exceeds $1\%$ in the time-interval shown. In other simulations we found that perturbation theory agrees better with full calculations if $\alpha$ is small, as is expected. Even for parameters with poorer agreement between full- and perturbative dynamics (for example larger $\alpha$ or closer to an avoided crossing), we typically still obtain the primary features which make these induced potentials interesting: (i) We have persistent dressed hopping, (ii) the excited state occupancy is suppressed.

Since we have included the dipole-dipole interaction in the \emph{unperturbed} part of the Hamiltonian \bref{Hunpert}, we do not a priori require it to be small compared to the detuning for the perturbation theory in \sref{dimer_vanVleck} to be valid. In practice we find that expression \bref{H_eff} works well even for $U_{ij} \sim \Delta_{s,p}$, as long as one stays clear of the avoided crossings at $U_{ij} = (\Delta_{s} + \Delta_{p})$. The large splitting of the eigenstates in the ground state manifold in the vicinity of the avoided crossing, see \frefp{spectrum_deltaeq}{b}, results in comparably fast population-oscillations. The regime where we can find dressed excitation transport on time scales shorter than the life-time of the multi atom state hence is typically entered when $U_{ij}$ approaches $\Delta_{s} + \Delta_{p}$, but has not quite reached it yet. The effective population-oscillation-period in \fref{comparison_PT_fullH} is $\sub{T}{ex} = \alpha^{-4}(1-\bar{U}_{12}^{2})/(2U_{12})=4.5$ms. Here $\alpha^{-4}=6.25\times 10^{6}$, and the correction factor is $(1-\bar{U}_{12}^{2})=0.13$ due to the vicinity of an avoided crossing. 

Let us reconsider the doubly excited state populations in \fref{N2_all}. For atomic distances where the system is far from all avoided crossings shown in \sref{surfaces}, the doubly excited states would be suppressed by $\alpha^4$. This gets modified near these avoided crossings. For the distance $R=6.5$ $\mu$m as, the system is fairly close to an avoided crossing, thus the doubly excited states are more strongly populated. 

To conclude this section, we would like to remark on the position degree of freedom of the atoms carrying the interaction. In this article we treat it classically, assuming atoms with a precisely defined separation $R$. In practice, each atom will have a position uncertainty, for example due to the zero-point motion in a harmonic well. The resulting distribution of interatomic distances $R$ can lead to a fast de-phasing of population oscillations such as shown in \fref{comparison_PT_fullH}. For the example shown, if we would assume a separation uncertainty of $\sigma_{R}=R/20$ the oscillations de-phase after $4$ cycles.

We however do not study de-phasing here, since it is not specific to the \emph{dressed} interactions. If we considered direct dipole-dipole excitation transport, between the states $\ket{sp}$ and $\ket{ps}$ only, it would also de-phase after $4$ cycles, which however would take place on a timescale of $10$ ns due to the by $\alpha^{-4}(1-\bar{U}_{12}^{2})$ stronger interaction.
Thus, whether or not de-phasing due to disorder poses a problem depends on the intended application of excitation transport and not on whether or not the interactions arise due to dressing.

In particular \rref{us:CI} discusses an exemplary application where it poses no problem, simulations presented in \rref{us:CI} \emph{fully include atomic positioning uncertainties}. Also 
in references \cite{cenap:motion,wuester:cradle,moebius:cradle} we study systems with interesting combined excitation transport and atomic motion, with interactions that could be interpreted as arising via the techniques considered here. Again, de-phasing effects do not pose a problem.

%%%%%%%%%%%%%%%%%%%%%%%%%%%%%%%%%%%
\subsection{Spontaneous decay and loss}
%%%%%%%%%%%%%%%%%%%%%%%%%%%%%%%%%%% 
\label{decay}

The use of chains of Rydberg atoms for random walks and other aggregate physics \cite{muelken:exciton:survival,wuester:cradle} is limited by spontaneous decay and other incoherent loss processes from the Rydberg states. When using a dressing scheme, these decay processes are reduced by construction since the atomic population in Rydberg states is kept small. Throughout the paper, \emph{except \sref{densitymatrix}}, we thus do not explicitly include loss in the presented dynamical calculations, but instead estimate the life-time of dressed Rydberg-aggregates based on their content of actual Rydberg population. For example the state $\ket{\tilde{\pi}_{1}}$ \eref{effective_states} will be assigned a total effective decay-rate $\sub{\gamma}{eff} =\alpha_{s}^2 \gamma_{s} + \alpha_{p}^2 \gamma_{p}$ for two atoms. This scaling of $\sub{\gamma}{eff}$ with the dressing parameters follows the same pattern as in vdW dressing \cite{nils:supersolids,pupillo:strongcorr:dressed}.

Most of our simulations treat excitation transport via $\nu=80$ Rydberg states of litihum, which have life-times $\tau_{s}=\gamma_{s}^{-1}=185.8$ $\mu$s and $\tau_{p}=\gamma_{p}^{-1}=315.8$ $\mu$s \cite{beterov:BBR}. This yields $\sub{\tau}{eff}=\sub{\gamma}{eff}^{-1}=290$ms for $\alpha=0.02$ and $N=2$. The corresponding exponential population decay is schematically indicated in \frefp{comparison_PT_fullH}{a}. 
Note, that simple estimates of the life-time of a dressed system using the relations of this section and the results of \sref{dimer_vanVleck} require excited state populations to be accurately described by perturbation theory. We have seen in \fref{N2_all} that this is the case, even fairly close to the avoided crossings in the spectrum.

%%%%%%%%%%%%%%%%%%%%%%%%%%%%%%%%%%%
\subsubsection{Density-matrix treatment}
\label{densitymatrix}

To ascertain that the simple estimates of loss effects discussed above are valid for our cases, we also calculated results such as shown in \fref{comparison_PT_fullH} with the inclusion of loss and spontaneous decay. Consider the density matrix 
\begin{eqnarray}
\hat{\rho}=\sum_{\bv{k},\bv{l}}^{'}  \rho_{\bv{k},\bv{l}} \ket{\bv{k}}\bra{\bv{l}},
\end{eqnarray}
with basis-states as defined in \eref{nbodybasis}. The prime on the summation symbol indicates that only states within the single exciton manifold (set $b_{1}$ in \aref{multi_exc}) are considered.

Let the time-evolution of $\hat{\rho}$ be given by the Master equation
\begin{eqnarray}
\label{master_eqn}
\hat{\rho}=-i \comm{\hat{H}}{\hat{\rho}}  + \sum_{ak}\bar{\gamma}_{k}{\cal D}[\hat{D}_{k}^{(a)},\hat{\rho}]  + \Gamma {\cal L}[\hat{L},\hat{\rho}],
\end{eqnarray}
where ${\cal D}$ is the Lindblad superoperator ${\cal D}[\hat{a},\hat{\rho}] =\hat{a}\hat{\rho}\hat{a}^{\dagger} - \hat{a}^{\dagger}\hat{a}\hat{\rho}/2 - \hat{\rho}\hat{a}^{\dagger}\hat{a}/2$. For $k=\{s,p\}$ define $l(k)=\{g,h\}$, then $\hat{D_{k}}^{(a)}$ acts like $\ket{l(k)}\bra{k}$ in the electronic space of atom $a$ and as unity elsewhere. These operators describe spontaneous decay.

The superoperator ${\cal L}$ has the form ${\cal L}[\hat{a},\hat{\rho}] = \hat{a} \hat{\rho}-\hat{\rho}\hat{a}^{\dagger}$, and the operator $\bra{\bv{k}}\hat{L}\ket{\bv{k}'}=\sub{n}{Ryd}(\bv{k})\delta_{\bv{k}\bv{k}'}/2i$. Here $\sub{n}{Ryd}(\bv{k})$ counts the number of excited atoms in the state $\ket{\bv{k}}$.
The purpose of ${\cal L}$ is to describe loss from the Rydberg states to states external to our model \cite{drake:atomicphysics}, for example to further Rydberg states. For simplicity we assumed an identical loss rate $\Gamma$ from $\ket{s}$ and $\ket{p}$.
\begin{figure}[htb]
\centering
\psfig{file= 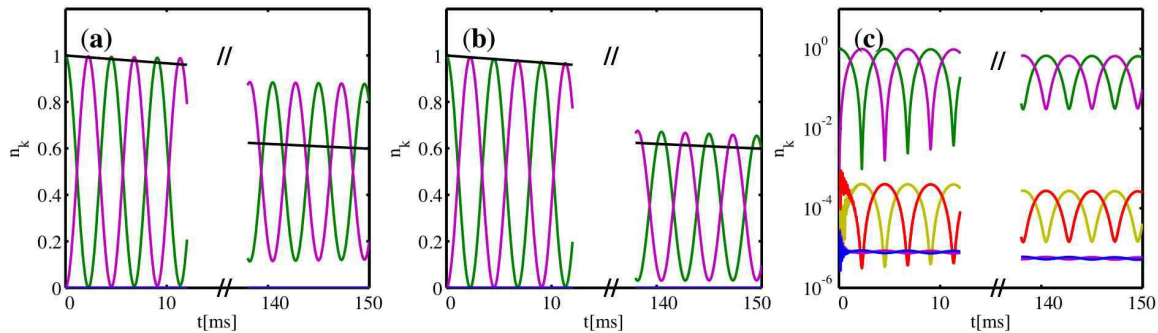,width=\columnwidth}
\caption{\label{incoherent_runs}The same scenario as \fref{comparison_PT_fullH}, but calculated with a density-matrix formalism including spontaneous decay and loss. (a) All decay processes return population to the ground states $\ket{g,h}$, case (i) described in the text. (b) Some decay processes loose population from our system, case (ii) described in the text. (c) The same as (b), but on a logarithmic scale. The black line in (a,b) is the function $p(t)=N(t)/N(0)=\exp{(-t/\sub{\tau}{eff})}$, with $\sub{\tau}{eff}=290$ ms, shown for orientation.
} 
\end{figure}

We consider two cases: (i) All the population loss from the Rydberg levels described by the effective decay rates $\gamma_{s,p}$ from \cite{beterov:BBR}, including black-body effects, ends up in the respective ground-states. Here we set $\Gamma=0$ and $\bar{\gamma}_{s,p}= \gamma_{s,p}$. This is a worst case scenario for our purposes, as all incoherent processes within our four state model de-cohere excitation transport. (ii) We split $\sub{\tau}{eff}$ from \cite{beterov:BBR} into the zero temperature component $\sub{\tau}{0}$ and the black-body induced component  $\sub{\tau}{bbr}$ via $\sub{\tau}{eff}^{-1}=\sub{\tau}{0}^{-1} + \sub{\tau}{bbr}^{-1}$. This yields $\sub{\tau}{0,s}=413.7$ $\mu$s, $\sub{\tau}{0,p}=1386$ $\mu$s, $\sub{\tau}{bbr,s}=337.3$ $\mu$s, $\sub{\tau}{bbr,p}=409.0$ $\mu$s. We then assume that population returns to their respective ground-states with a rate $\bar{\gamma}_{s,p}=(\sub{\tau}{0,(s,p)})^{-1}$, while black-body redistributed population is lost, either due to ionisation or because it leaves the single exciton manifold\footnote{For example through a cascaded decay like $\ket{gp}\rightarrow \cdots \rightarrow \ket{gg}$.}. We thus set $\Gamma=\sub{\tau}{bbr,s}^{-1}$.

These calculations, shown in \fref{incoherent_runs}, verify that loss out of the system from the Rydberg states $\ket{s}$ and $\ket{p}$ has indeed no other consequences than an overall population decay. If spontaneous decay in the channels $\ket{s}\rightarrow\ket{g}$ and $\ket{p}\rightarrow\ket{h}$ is included, population oscillations additionally show a de-phasing. Neither effect is dramatic on the time-scales considered here.

%%%%%%%%%%%%%%%%%%%%%%%%%%%%%%%%%%%
\subsection{Dressing in singly excited state manifolds}
%%%%%%%%%%%%%%%%%%%%%%%%%%%%%%%%%%%
\label{singleexc}

In \sref{surfaces} we have pointed out that dressing induced dipole-dipole interactions can be created in the ground-state manifold (between the states $\ket{\tilde{gh}}$, $\ket{\tilde{hg}}$) or in two different excited state manifolds (between the states $\ket{\tilde{sh}}$, $\ket{\tilde{hs}}$ or $\ket{\tilde{gp}}$, $\ket{\tilde{pg}}$). The latter requires significantly different detunings $\Delta_{s}$ and  $\Delta_{p}$ (such like $\Delta_{s}=-\Delta_{p}$) to energetically separate the two different singly-excited state manifolds. We have explicitly verified population-oscillations for example between the states $\ket{hs}$ and $\ket{sh}$.

To know whether the ground or singly-excited state manifold would be more useful for a specific application it is important to consider the life-time of the dressed multi-atom state: The smaller $\alpha$, the longer the life-time, but also the longer the effective hopping period $\sub{T}{ex}\sim 1/(2\sub{U}{eff})$ required for a complete population-oscillation that transfers the excitation from one atom in the dimer to its counter-part and back.
Here, the interaction $\sub{U}{eff}$ denotes the off-diagonal entry in the effective Hamiltonian of the corresponding manifold, e.g.~$\tilde{U}_{21}(R)$ in \eref{ER} for the ground-state manifold.

Here we estimate the life-time of dressed states more roughly than described in \sref{decay}, assuming that ground-state atoms ($\ket{g}$, $\ket{h}$) do not decay and excited state atoms ($\ket{s}$, $\ket{p}$) decay with the same rate $\gamma=1/\tau_{0}$. For a system of $N$ atoms in various states we then again add the decay rates. 

\begin{table}
\begin{center}
\begin{tabular}{|c|ccc|}
\cline{1-4}
\Big. 
state      &   $\sub{\tau}{eff}$   & $\sub{U}{eff}$  &$\beta=\sub{\tau}{eff}/\sub{T}{ex}$ \\
\cline{1-4}
\Bigg. $\ket{\tilde{sss\cdots sp}}$ &  $\frac{\tau_{0}}{N}$ & $U$  &$2\tau_{0} U\frac{1}{N} \equiv \beta_{0}\frac{1}{N}$\\
\Bigg. $\ket{\tilde{ggg\cdots gp}}$ &  $\frac{\tau_{0}}{1 + (N-1)\alpha^2 }$ &  $\alpha^2 U$ &  $\beta_{0}\frac{1}{N-1 + \alpha^{-2}}$ \\
\Bigg. $\ket{\tilde{ggg\cdots gh}}$ & $\frac{\tau_{0}}{N\alpha^2}$   &  $\alpha^4 U$  &$\beta_{0}\frac{\alpha^2}{N}$\\
\cline{1-4}
\end{tabular}
\end{center}
\caption{Lifetimes and interaction strengths of various dressed $N$-atom states: From top to bottom in the excited-state manifold, singly-excited-state manifold and ground-state manifold.
\label{lifetimes}}
\end{table}
For three different $N$-atom states, the life-times determined in this way and the ratio of life-time and hopping period $\beta=\sub{\tau}{eff}/\sub{T}{ex}$ are listed in \tref{lifetimes}. As long as $\alpha^{-2}\lesssim N$, the ratio $\beta$ for the singly-excited state manifold can be larger than for the ground-state manifold.

Whether dipole-dipole interactions induced through dressing are advantageous over direct dipole-dipole interactions among Rydberg states, strongly depends on the problem at hand, as we discuss now for three examples. 

\ssection{(i) Exciton migration} For the migration of a single or multiple excitons on a rigid chain, as studied in \cite{muelken:exciton:survival}, the ratio $\beta$ introduced above gives a direct measure of how many sites the excitation can traverse within the life-time of the whole chain. Evidently this measure is never improved by the present dressing technique, whose advantage for this scenario thus only lies in reduced ionization probabilites and simpler atom trapping. In \sref{chain} we show an exemplary case where the overall life time in the presence of dressing is long enough to exploit these features.

\ssection{(ii) Adiabatic transport} There are dynamical scenarios whose time-scale of interest is not directly given by $\sub{T}{ex}$, such as the adiabatic entanglement and excitation transport that we reported in \cite{wuester:cradle}. It can be seen that the ratio of the time-scale of transport dynamics to spontaneous life-time of the chain can be improved by factors of about 2 by working in the dressed singly-excited-state manifold.   

\ssection{(iii) Ring aggregates} In \sref{trimer} we consider systems of three dipole-dipole interacting atoms trapped in effectively one-dimension on a ring. For such a construction dressing can be highly beneficial.

%%%%%%%%%%%%%%%%%%%%%%%%%%%%%%%%%%%
%%%%%%%%%%%%%%%%%%%%%%%%%%%%%%%%%%%
\section{Applications} 
%%%%%%%%%%%%%%%%%%%%%%%%%%%%%%%%%%%
%%%%%%%%%%%%%%%%%%%%%%%%%%%%%%%%%%%
\label{applications}

After choosing the dimer as the simplest case to work out the basic details of dressed dipole-dipole interactions, we now briefly present two exemplary applications of our results to larger systems. 
First, in \sref{trimer}, we consider a dressed ring-trimer ($N=3$ atoms confined on a ring), a system of interest due to conical intersections of the adiabatic energy surfaces \cite{us:CI}.
Then, in \sref{chain}, we examine a dressed exciton on an atomic chain with $N=5$. 

%%%%%%%%%%%%%%%%%%%%%%%%%%%%%%%%%%%
\subsection{Flexible Rydberg ring-trimer} 
%%%%%%%%%%%%%%%%%%%%%%%%%%%%%%%%%%%
\label{trimer}

In \rref{us:CI} we show that circular Rydberg trimers, consisting of three Rydberg atoms trapped in one dimension on a ring, exhibit interesting quantum dynamics near conical intersections (CIs). For practical realisations, however, dipole-dipole forces are typically \emph{too large} for ring confinement. In this case the reduction of interaction strengths, which limits the parameter range where dressing is beneficial for  exciton migration, turns into a benefit. As dressing also increases the system's life-time, it greatly facilitates the practical creation of dipole-dipole ring trimers.

\begin{figure}[htb]
\psfig{file= 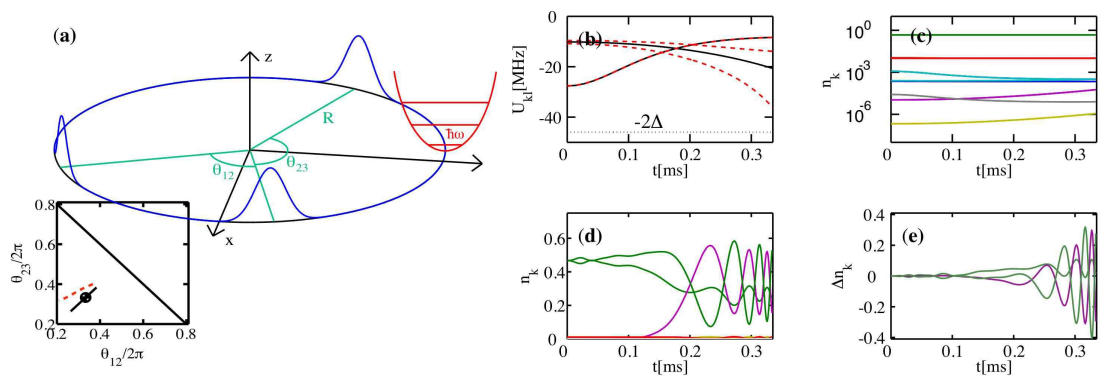,width=\columnwidth}
\caption{\label{trimer:plots} Adiabatic dressing dynamics according to \eref{Hamil_dressed} for temporally varying bare interaction strength $U_{kl}(t)$. (a) Geometric arrangement of dressed Rydberg ring trimer as discussed in \rref{us:CI}. The interaction strength is varied according to two different (classical) trajectories of motion of the ring-trimer as shown in the inset. For the solid path, symmetry forces the eigenstates of \eref{Hamil_dressed} to remain constant in time despite changes in $U_{kl}(t)$, this corresponds to straight crossing of the conical intersection of the (dressed) energy surfaces. For the red dashed path, dressed eigenstates change in time and are being adiabatically followed, corresponding to motion on the same energy surface in \rref{us:CI}. (b) Bare interaction strengths $U_{kl}(t)$ for the two paths, using the same coding. For comparison $-2\Delta $ is also shown, where $\Delta=24$MHz is the excited state detuning. (c) Logarithmic plot of populations $n_{k}=|c_{k}|^2$, for the solid path in the inset of (a), using the same assignment of colours to states excitation number as in \fref{N2_all}. Grey indicates triply excited states. (d) Populations $n_{k}=|c_{k}|^2$ for the dashed path in (b), using a normal scale. (e) Difference between the exact populations in (d) and the expected dynamics according to the fourth order effective perturbative Hamiltonian \eref{HeffN}. }
\end{figure}

Consider three atoms tightly confined in one dimension on a ring, as sketched in \frefp{trimer:plots}{a}. Achieving such confinement is the first point in this section that is simpler for dressed ground state atoms than bare Rydberg atoms. On the ring, the bare interaction between atoms $U_{kl}$ is fully determined by the two relative angles $\theta_{12}$ and $\theta_{23}$ shown in the sketch. Adiabatic energies, defined by \eref{eigenvalue_eqn}, then form 2D surfaces $E_{k}(\theta_{12},\theta_{23})$. In the absence of laser-couplings, the three surfaces spanned by $\ket{pss}$, $\ket{sps}$, $\ket{ssp}$ exhibit a conical intersection \cite{yarkony2001conical}, where two of them become exactly degenerate, at $\theta_{12}=\theta_{23}=2\pi/3$. In the presence of laser-couplings, this feature is imprinted also onto surfaces spanned predominantly by ground states $\ket{hgg}$, $\ket{ghg}$, $\ket{ggh}$.
Conical intersections are of great interest in particular in chemical physics, as they strongly affect the outcome of photo-chemical reactions \cite{yarkony2001conical}. In cold atomic gases of Rydberg dressed atoms, they result in interesting non-adiabatic and geometric phase effects \cite{us:CI}. For a detailed description of this system and the concept of conical intersections we refer to \rref{us:CI} and references therein.

For the purpose of the present article, it is more important to point out that the techniques worked out here are highly beneficial for the practical realisation of this kind of ultra cold conical intersections.
This conclusion is reached through an extensive but technical survey of parameter space. We defer full details to a specialised publication, and present here only the core points:

In order to realise the scenario sketched in \frefp{trimer:plots}{a}, initially at least three conditions have to be met:
\\ \noi
{\it A. The dipole-dipole interaction energy should not exceed the transverse oscillator spacing $\omega_{\perp}$ to ensure one-dimensional dynamics.}  Our interest is in dynamics near the conical intersection, where the atoms form an equilateral triangle and are separated by $d=\sqrt{3}R$, where $R$ is the ring radius. Comparing the (possibly dressed) interaction strength with the strength of the ring trap, $\omega_{\perp}$, we have $\mu^{2}/d^3\alpha^4<\omega_{\perp}$, thus preventing either too small ring radii $R$ or to large interaction strength $\mu$ and hence principal quantum numbers $\nu$.
\\ \noi
{\it B. The motion should be adiabatic, except close to the conical intersection.} We can estimate the order of magnitude of the CI transit time $T_{CI}$ from the time-scale set by the classical equations of 
motion $\ddot\theta=-(\partial V(\theta )/\partial \theta)/(2MR^2)$, which holds for a dimer of angular separation $\theta$. This time scale is $T_{CI}=2^{3/4} \sqrt{M} R^{5/2}/(\mu\alpha^{2})$. Adiabatic dynamics must fulfill $T_{CI} \gg \sub{T}{ex}$, with $\sub{T}{ex}$ as in \sref{singleexc}. 
\\ \noi
{\it C. The life-time of the dressed three-atom system should be longer than the duration of the motion,}  
thus $\sub{\tau}{eff}> T_{CI}$.
\\ \noi
Attempting to fulfill the inequalities (A-C) for various atomic species leads to two main conclusions: (i) Lighter atoms are favourable. (ii) Without dressing ($\alpha=1$), principal quantum numbers $\nu$ would be too high to avoid strong effects of black-body radiation. If we consider a dressed system however, one more constraint comes into play:
\\ \noi
{\it D. The \emph{bare} dipole-dipole shift without dressing must be smaller than the doubly excited state detuning $\Delta_{s}+\Delta_{p}$, at the closest approach of the atoms.} This is to stay clear of avoided crossings such as seen in \fref{spectrum_deltaeq}.

A set of parameters where these conditions are met is: $\nu=100$, $R=9.8$ $\mu$m, $\Delta_{s}=\Delta_{p}=24$ MHz and $\alpha_{s}=\alpha_{p}=0.15$
\footnote{
These are the parameters employed in \cite{us:CI}, in that article the detuning and $\sub{V}{bare}(d^{*})$ were given as angular frequency $\Delta_{\omega}=2\pi \times \Delta$.
}. 
Now we consider the dynamical evolution of an eigenstate of \eref{Hamil_dressed}, when the interaction parameters $U_{kl}=U_{kl}(t)$ are varied in time in a manner representative for near CI dynamics. To this end we extracted the trajectory $\btheta(t)=\langle \bv{\btheta} \rangle$ from full quantum mechanical solutions of the time-dependent Schr\"odinger equation in \rref{us:CI}\footnote{These simulations only contained the three effective ground-state energy surfaces.}, shown in the inset of \frefp{trimer:plots}{a} as dashed line. A second trajectory is taken slightly offset (dotted line). We find that the quantum state dynamics in both cases follows $\sub{H}{eff}$ (see equation \eref{HeffN}) well, justifying the simplified model used in \cite{us:CI}. Since the energy-spacing $E_{pd}$ (see \fref{fig:sketch}) for $\nu=100$ is $E_{pd}=298$ MHz, we could have chosen a much larger detuning, yielding even better agreement between the perturbative and exact solutions. 

As we have seen in this section, dressing is threefold beneficial for atom trapping in the construction of conical intersections in dressed Rydberg ring trimers: (i) It greatly widens the available scope for quasi-1D trapping of the atoms, (ii) reduces the involved interaction strength such that trapping can be realistically considered in the first place and (iii) simultaneously extends the available life-times.

%%%%%%%%%%%%%%%%%%%%%%%%%%%%%%%%%%%
\subsection{Dressed excitation transport on long atomic chains} 
%%%%%%%%%%%%%%%%%%%%%%%%%%%%%%%%%%%
\label{chain}

We discuss results for a linear chain of $N=5$ equidistant atoms in the following, considering the same physical situation as in \sref{checks}, hence the spacing of atoms in the chain is also $\Delta R=6.5\mu$m. We create the initial state adiabatically as in \sref{checks}.

For this section only, we will define our excitation to be the state $\ket{g}$, such that e.g.~$\ket{\pi_{1}}=\ket{ghhhh}$. The corresponding dressed states have a longer life-time for larger chains.

\begin{figure}[htb]
\psfig{file= 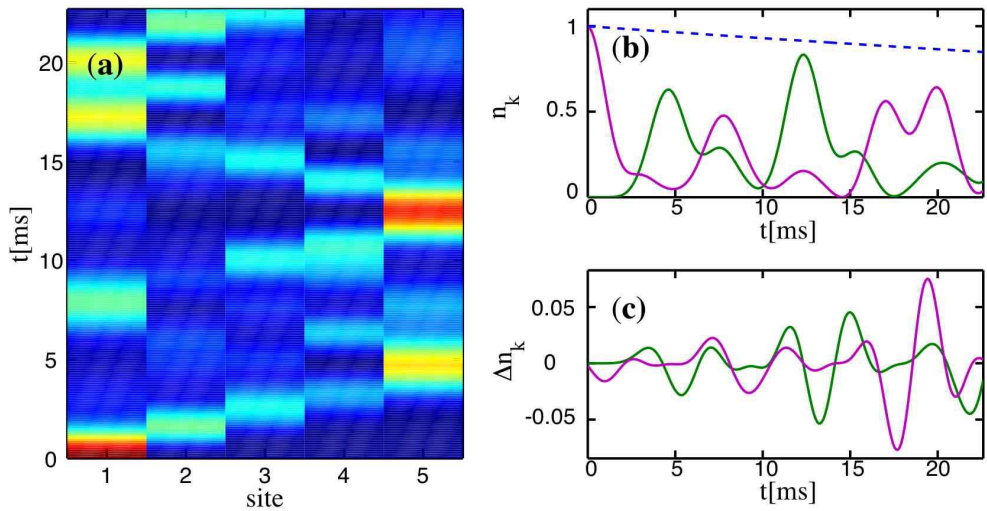,width=\columnwidth}
\caption{\label{N5_chain}(a) Dressing-induced excitation transport in a chain of $5$ atoms, according to the Hamiltonian \bref{HeffN}. Visualized is the population of states $\ket{\pi_{n}}$. (b) Quantitative details of the same data as in (a). Sites $1$ and $5$ are shown in (magenta, green) respectively, the approximate population loss is shown as blue dashed line. (c) Absolute difference between the evolution according to the fourth order effective Hamiltonian \bref{HeffN} and the full $5$-body Hamiltonian \bref{Hamil_dressed}. 
}
\end{figure}
The dressing-induced excitation transport and a comparison of exact and perturbative evolution is shown in \fref{N5_chain}. It can be seen that the excitation is transported over many sites of the chain on time-scales shorter than the expected spontaneous life-time of the dressed $5$-atom state (about $\sub{\tau}{eff}=[\alpha^{2}((N-1)/\tau_{p} + 1/\tau_{s}]^{-1}=0.14$ s). Due to the dressing, the time-scale on which the interaction transports the excitation has been increased by a factor of $\alpha^{-4}(1-\bar{U_{12}})=8.4\times10^5$, compared to the direct use of excited Rydberg states. 

 In \sref{singleexc} we argued that dressing typically does not favourably alter the possible number of excitation hops, given by the ratio $\beta=\sub{\tau}{eff}/\sub{T}{ex}$ of life-time and hopping period. Nonetheless the results of this section show that dressed excitation migration within the available life-time is possible. Despite the detrimental effect on $\beta$, the dressing still enables time-dependent control of the transport, and, as in the previous section, simplifies trapping efforts.

%%%%%%%%%%%%%%%%%%%%%%%%%%%%%%%%%%%
%%%%%%%%%%%%%%%%%%%%%%%%%%%%%%%%%%%
\section{Conclusions and Outlook} 
%%%%%%%%%%%%%%%%%%%%%%%%%%%%%%%%%%%
%%%%%%%%%%%%%%%%%%%%%%%%%%%%%%%%%%%
\label{conclusion}

We have shown that the dressing of alkali atoms in the ground-state with Rydberg-excitations can yield dipole-dipole interactions with excitation-transfer for the ground-state atoms. This generalises existing results for dressing with Rydberg van der Waals interactions, that do not entail excitation transport. The scheme proposed here is an adaptation of a similar method, introduced in the context of trapped Rydberg ions \cite{mueller:iondressing}, to neutral atomic systems. It makes use of two effective laser couplings to Rydberg-states, instead of the one required for van der Waals dressing. 

We find that the dressed dipole-dipole transition matrix element between ground state atoms scales like $\alpha^4$ with the dressing parameter $\alpha$, while the life-time of the dressed state increases like $\alpha^{-2}$. Dressed vdW interactions show the same scaling behaviour. We demonstrate in \rref{us:CI} that this re-arrangement of life-times and interaction time-scales allows the study of conical intersections using Rydberg-dressed atoms on a ring. Beyond this, the dressing scheme will enable a larger variety of trapping techniqes for the atoms and reduce the likelihood of ionisation. As further example of the flexibility afforded by the dressing, we show that it is possible to induce effective transport of a single Rydberg excitation within a chain of ground state atoms. The amplitude of this process
scales like $\alpha^2$.

Whether the effective interactions discussed in this article are more useful than the direct use of dipole-dipole interactions depends on details of the situation one studies. In \rref{us:CI} we provide an exemplary scenario where they can be highly beneficial.

%%%%%%%%%%%%%%%%%%%%%%%%%%%%%%%%%%%%%%%%%%%%%%%%%%%%%%%%%%%%%%%%%
\ack{We thank Igor Lesanovsky, Markus M{\"u}ller, Rejish Nath, Thomas Pohl and Sevilay Sevin{\c c}li for helpful comments.}
%%%%%%%%%%%%%%%%%%%%%%%%%%%%%%%%%%%%%%%%%%%%%%%%%%%%%%%%%%%%%%%%%

%%%%%%%%%%%%%%%%%%%%%%%%%%%%%%%%%%%
%%%%%%%%%%%%%%%%%%%%%%%%%%%%%%%%%%%
\appendix
%%%%%%%%%%%%%%%%%%%%%%%%%%%%%%%%%%%
%%%%%%%%%%%%%%%%%%%%%%%%%%%%%%%%%%%
\section{Atom-light interactions} 
%%%%%%%%%%%%%%%%%%%%%%%%%%%%%%%%%%%
%%%%%%%%%%%%%%%%%%%%%%%%%%%%%%%%%%%
\label{rotwave}

In this section, we derive the rotating-wave Hamiltonian \eref{Hamil_dressed} from more fundamental expressions for atom-laser interactions, reviewing text-book material, see e.g.~\cite{book:scully}. 
The Hamiltonian for an atom in the presence of an electric field is
\begin{eqnarray}
\label{Hamil_Efield}
\hat{H'}  = \hat{W}_{0} - e \hat{\bv{r}}\cdot \bv{E}(\bv{r}_{0},t).
\end{eqnarray}
This expression is valid in the dipole approximation $\bv{k}\cdot \bv{r}\ll 1$, when the wavelength of the incident light is much larger than typical atomic distances. The operator $\hat{W}_{0}$ is the full atomic Hamiltonian in the absence of any external field, $e$ is the electron charge, $\hat{\bv{r}}$ the position operator and ${\bv{E}}(\bv{r}_{0},t)$ the electric field at the atomic nucleus, which is treated classically. 

Consider for the moment a single atom only, with the four relevant states $\ket{g}$, $\ket{h}$, $\ket{s}$, $\ket{p}$ and a level scheme as shown in \frefp{fig:sketch}{a}. These states are eigenvectors of the free atomic Hamiltonian $\hat{W}_{0}$, thus $\hat{W}_{0}\ket{k}=\hbar E_{k}$ with $k\in\{g,h,s,p\}$. We add two 
\footnote{For \emph{each} of the two transitions that we require, we only consider a \emph{single} light-field instead of the multiple lasers that would typically be employed for a multi-photon transition. A more detailed treatment does not add qualitatively new features.}
coupling fields $\bv{E}=\bv{E}_{s}+\bv{E}_{p}$, with $\bv{E}_{s,p}={\cal E}_{s,p}\hat{\bv{z}} \cos{(\nu_{s,p} t)}\equiv {\cal E}_{s,p}(t)\hat{\bv{z}}$. Here $\hat{\bv{z}}$ is a unit vector in the $z$-direction. For any two states $\ket{a}$, $\ket{b}$ we define matrix elements $\hat{\mu}_{ab}=e\bra{a}\hat{z}\ket{b}$. A general quantum state in the chosen subspace is $\ket{\Psi(t)}=\sum_{n\in\{g,h,s,p\}}c_{n} \ket{n}$, with equation of motion
\begin{eqnarray}
\label{effective_equation1}
\!\!\!\!\!\!\!\!\!\!\!\!\!\!\!\!\!\!\!\!\!\!\!\!\!\!\!\!\!\!\!\!\!\!\!\!\!\!\!\!\!\!
i\frac{\partial}{\partial t} 
\left[  
\begin{array}{c}
c_{g}\\
c_{s}\\
c_{h}\\
c_{p}\\
\end{array} 
\right]  
&=
\left[ \begin{array}{ccccc}
E_{g} & -\mu_{gs} {\cal E}_{s}(t) & 0 & 0 \\
-\mu_{gs}^{*} {\cal E}_{s}(t) &  E_{s} & 0 & 0 \\
0 & 0 & E_{h} & -\mu_{hp} {\cal E}_{p}(t)\\
0 & 0 & -\mu_{hp}^{*} {\cal E}_{p}(t) &  E_{p} \\
\end{array} 
\right]  
\left[  
\begin{array}{c}
c_{g}\\
c_{s}\\
c_{h}\\
c_{p}\\
\end{array} 
\right].
\end{eqnarray}
Here we have used atomic units.
Now we change the variables $c_{m}$ of the coefficient vector to $d_{m}=c_{m}e^{i\omega_{m} t}$, where $\omega_{m}$ are arbitrary constants. If the coefficient vector $\bv{c}$ evolved according to $i\pdiff{}{t}\bv{c}=M \bv{c}$ for some matrix $M$, $\bv{d}$ evolves according to $i\pdiff{}{t}\bv{d}=M' \bv{d}$ with $M' = E M E^* - \Omega$, where $E$ and $\Omega$ are diagonal matrices with elements $E_{mm}=e^{i\omega_{m} t}$ and $\Omega_{mm}=\omega_{m}$.

Using the specific transformation vector $\bomega=(E_{g},E_{s},E_{h},E_{p})^{T}$, we obtain
\begin{eqnarray}
\label{effective_equation2}
\!\!\!\!\!\!\!\!\!\!\!\!\!\!\!\!\!\!\!\!\!\!\!\!\!\!\!\!\!\!\!\!\!\!\!\!\!\!\!\!\!\!
i\frac{\partial}{\partial t} 
\left[  
\begin{array}{c}
d_{g}\\
d_{s}\\
d_{h}\\
d_{p}\\
\end{array} 
\right]  
&=
\left[
 \begin{array}{c}
0  \\
-\frac{1}{2}\mu_{gs}^{*} {\cal E}_{s}  f_{-}(\omega_{s},\nu_{s})  \\
0  \\
0  \\
\end{array} 
\!\!\!\!\!\!\!\!\!\!\!\!\!\!\!\!\!
 \begin{array}{c}
-\frac{1}{2}\mu_{gs} {\cal E}_{s}  f_{+}(\omega_{s},\nu_{s})   \\
0 \\
0  \\
0  \\
\end{array} 
\!\!\!\!\!\!\!\!\!\!\!\!\!\!\!\!\!
 \begin{array}{c}
0  \\
0  \\
0  \\
-\frac{1}{2}\mu_{hp}^{*} {\cal E}_{p}   f_{-}(\omega_{p},\nu_{p}) \\
\end{array} 
\!\!\!\!\!\!\!\!\!\!\!\!\!\!\!\!\!
 \begin{array}{c}
0  \\
0  \\
 -\frac{1}{2}\mu_{hp} {\cal E}_{p} f_{+}(\omega_{p},\nu_{p})  \\
0  \\
\end{array} 
\right]  
%\left[ \begin{array}{cccc}
%0 & -\frac{1}{2}\mu_{gs} {\cal E}_{s} e^{-i(\omega_{s}\pm \nu_{s})t} & 0 & 0 \\ 
%-\frac{1}{2}\mu_{gs}^{*} {\cal E}_{s}  e^{i(\omega_{s}\pm \nu_{s})t}&  0 & 0 & 0 \\
%0 & 0 & 0 & -\frac{1}{2}\mu_{hp} {\cal E}_{p} e^{-i(\omega_{p}\pm \nu_{p})t}\\
%0 & 0 & -\frac{1}{2}\mu_{hp}^{*} {\cal E}_{p}  e^{i(\omega_{p}\pm \nu_{p})t}&  0 \\
%\end{array} 
%\right]  
\left[  
\begin{array}{c}
d_{g}\\
d_{s}\\
d_{h}\\
d_{p}\\
\end{array} 
\right],
\end{eqnarray}
 where $\omega_{s}=E_{s} -E_{g}$, $\omega_{p}=E_{p} -E_{h}$ and $f_{\pm}(\omega,\nu)$ stands for $e^{\pm i(\omega+ \nu)t} + e^{\pm  i(\omega- \nu)t}$. Under the \emph{rotating-wave approximation} all terms $e^{\pm i(\omega_{s,p}+ \nu_{s,p})t}$ are dropped, as they are oscillating too rapidly to have any effect. We also introduce the detunings $\Delta_{s,p}= \nu_{s,p}-\omega_{s,p}$ and Rabi-frequencies $\Omega_{s}= - \mu_{gs} {\cal E}_{s}$ and $\Omega_{p}= - \mu_{hp} {\cal E}_{p}$.
 
Finally we perform one further variable-change, using $\bomega =(0, \Delta_s,0,\Delta_p)^{T}$ to reach
\begin{eqnarray}
\label{effective_equation3}
i\frac{\partial}{\partial t} 
\left[  
\begin{array}{c}
d'_{g}\\
d'_{s}\\
d'_{h}\\
d'_{p}\\
\end{array} 
\right]  
&=
\left[ \begin{array}{ccccc}
0 & \frac{\Omega_{s}}{2} & 0 & 0 \\
\frac{\Omega_{s}^{*}}{2}&  -\Delta_{s} & 0 & 0 \\
0 & 0 & 0 & \frac{\Omega_{p}}{2} \\
0 & 0 & \frac{\Omega_{p}^{*}}{2} &   -\Delta_{p} \\
\end{array} 
\right]  
\left[  
\begin{array}{c}
d'_{g}\\
d'_{s}\\
d'_{h}\\
d'_{p}\\
\end{array} 
\right].
\end{eqnarray}
The Hamiltonian in \bref{effective_equation3} is the basic building block of \eref{Hamil_dressed}, as long as only a single atom is concerned. Even for more than one atom and with the inclusion of
dipole-dipole interactions, the above procedure can be followed. Instead of the single atom energies in the first transformation vector $\bomega $, we would employ many-atom energies. Since the dipole-dipole interaction only couples energetically degenerate states like $\ket{sp}$ to $\ket{ps}$, all complex phase-factors can finally be eliminated to arrive at a many-body version of \eref{effective_equation3}, which is \eref{Hamil_dressed}. 

%%%%%%%%%%%%%%%%%%%%%%%%%%%%%%%%%%%
%%%%%%%%%%%%%%%%%%%%%%%%%%%%%%%%%%%
\section{Lithium energy levels} 
%%%%%%%%%%%%%%%%%%%%%%%%%%%%%%%%%%%
%%%%%%%%%%%%%%%%%%%%%%%%%%%%%%%%%%%
\label{full:levels}

\begin{figure}[htb]
\psfig{file= 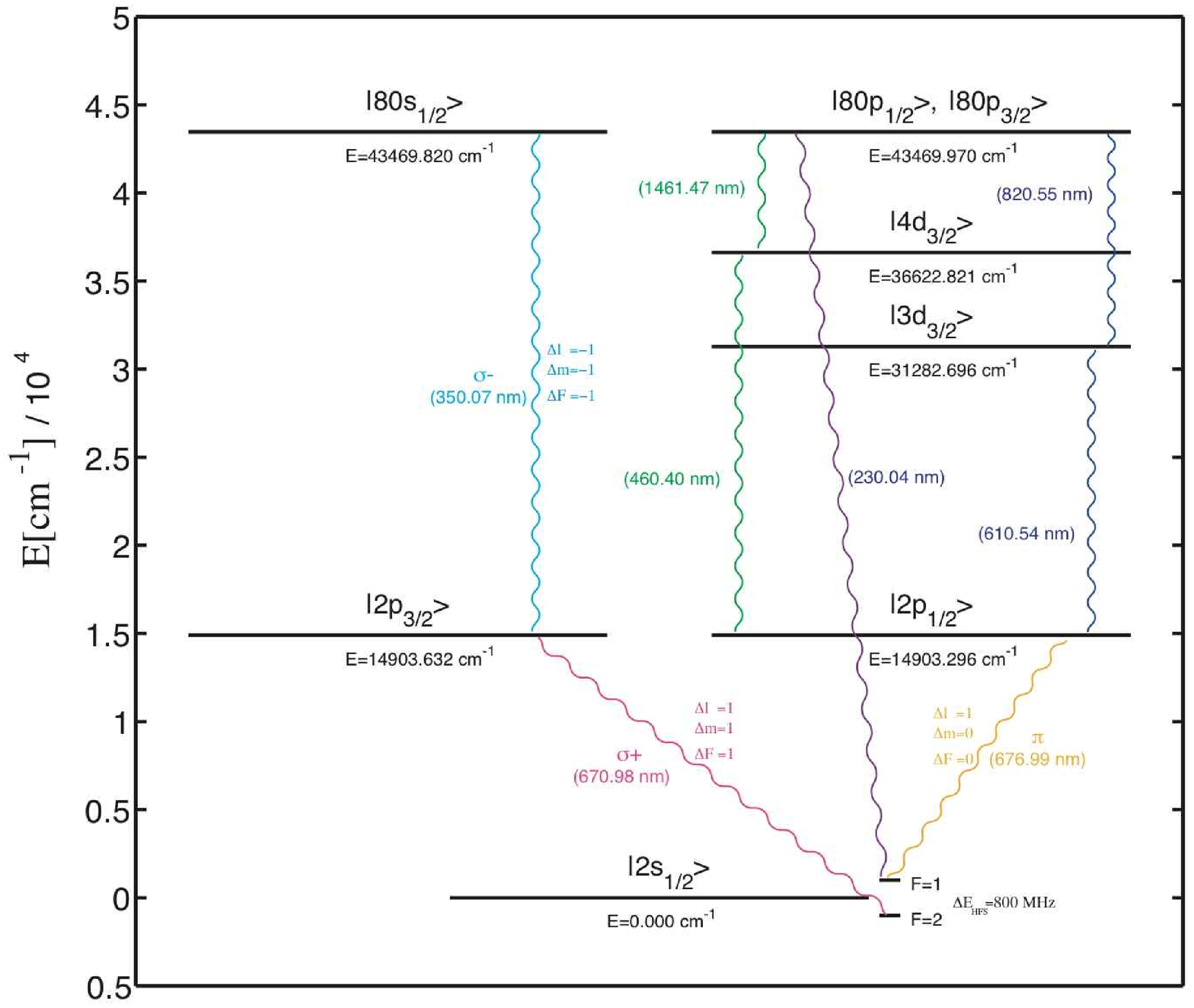,width=0.75\columnwidth}
\caption{\label{full:level:fig} Several alternative transition path-ways to realise the overall effective coupling between the ground- and Rydberg states. The $\ket{g}\leftrightarrow\ket{s}$ coupling proceeds via a standard two-photon transition, while the $\ket{h}\leftrightarrow\ket{p}$ coupling requires either a UV-transition \cite{tong:blockade} or a three-photon transition \cite{anwarulhaq:threephoton}, both of which are more cumbersome.
This diagram is intended as a rough guide only, hence fine- and hyperfine structure are omitted for all excited states. Intermediate state energies are taken from \cite{radziemski:lilevels}, Rydberg energies calculated with simple quantum defect theory as in \cite{stevens:Lidefects}.
}
\end{figure}

In practice the scheme displayed in \fref{fig:sketch} is complicated by selection rules and transitions are constrained by the availability of laser sources.
The coupling between the states $\ket{g}\leftrightarrow\ket{s}$ can be realised with now broadly established two-photon excitation schemes. These transitions are typically near resonant with some auxiliary middle level, such as $\ket{2p_{3/2}}$ in \fref{full:level:fig}, in order to enhance transition amplitudes. For the right choice of parameters, coherent coupling between $\ket{g}=\ket{2s_{1/2}, F=1}$ and  $\ket{s}=\ket{80s_{1/2}, F=1}$ is achieved, see e.g.~\cite{reetzlam:excitationtrapping}. For the coupling between the states $\ket{h}\leftrightarrow\ket{p}$ two-photon transitions are forbidden due to the selection rule $\Delta l=1$.

This necessitates one- or three photon coupling schemes, examples for which are included in \fref{full:level:fig}.
Let us briefly discuss both options: \\
\ssection{(i) Single photon UV-transition} Direct transitions from ground- to Rydberg levels \cite{tong:blockade} suffer from extremely small dipole-transition matrix elements.
These scale like $\nu^{-3/2}$ with the principal quantum number \cite{book:gallagher}. From the matrix element $|\bra{13p_{3/2}}\hat{r}\ket{2s_{1/2}}|=0.036a_{0}$ \cite{NIST:database}
\footnote{
We calculate matrix elements ${\cal M}$ from oscillator strengths $f_{ik}$ and transition energies ${\Delta}E$ using ${\cal M}=\sqrt{3e^{2}\hbar^{2} f_{ik}/(2 m_{e} {\Delta}E)}$, $m_{e}$ is the electron mass.
}, we extrapolate ${\cal M}=|\bra{80p_{3/2}}\hat{r}\ket{2s_{1/2}}|=0.0024 a_{0}$, where $a_{0}$ is the Bohr radius. Assuming $P=0.09$ mW of laser-power, focussed to a waist $w=10$ $\mu$m, one achieves a Rabi-frequency $\Omega=4$ MHz as employed in this article. The intensity at the focus is $I=2P/(\pi w^{2})$, resulting in a Rabi-frequency $\Omega={\cal M} e {\cal E}/\hbar$, where ${\cal E}=\sqrt{2 I/(c \epsilon_{0})}$ is the electric field, and $e$, $c$ as well as $\epsilon_{0}$ fundamental constants. 
Laser light at the UV wavelength is typically created through higher harmonic generation, reducing flexibility and making it more challenging to 
achieve the required power.
\\
\ssection{(ii) Three photon transition}
One would probably employ three different transitions, as indicated in \fref{full:level:fig}, exploiting near resonant auxiliary levels to enhance the coupling amplitude. In principle it would also be possible to utilise three photons stemming all from the same laser. For Li such three-photon, single colour, schemes still can proceed near resonant with the intermediate $\ket{2p_{1/2}}$ state~\cite{anwarulhaq:threephoton}.
Consider for example the excitation chain sketched in yellow and dark blue in \fref{full:level:fig}, and denote the Rabi-frequencies and detunings of the three transition as $\Omega_{k}$, $\Delta_{k}$ with $k=1,2,3$ from lower to higher energies. Since both intermediate states decay on a timescale of $10$ ns, they have to be far detuned with $\alpha_1=\Omega_{1}/(2\Delta_{1})$ and $\alpha_2=\sub{\Omega}{2,eff}/(2(\Delta_{1}+\Delta_{2}))$ of the order of $\alpha_{k}\lesssim 10^{-3}$ for $k=1,2$. Here $\sub{\Omega}{2,eff}=\alpha_1\Omega_{2}$. This makes sure that the effective decay rate of the intermediate levels is at least not larger than that due to the Rydberg state.

If we now consider the effective three-photon Rabi frequency:
\begin{eqnarray}
\label{effective_threephoton_Rabifreq}
\sub{\Omega}{eff}=\frac{\Omega_{1}\Omega_{2}\Omega_{3}}{4\Delta_{1}(\Delta_{1}+\Delta_{2})}=\alpha_1\alpha_2\Omega_{3},
\end{eqnarray}
we would require $\Omega_{3}\gtrsim4\times 10^{3}$ GHz to reach $\sub{\Omega}{eff}=4$ MHz as employed here, rather hard to achieve on the weak Rydberg transition. There may though still be applications where much smaller effective Rabi-frequencies $\sub{\Omega}{p}$ are sufficient and the three-photon scheme has advantages over the UV transition.

%%%%%%%%%%%%%%%%%%%%%%%%%%%%%%%%%%%
%%%%%%%%%%%%%%%%%%%%%%%%%%%%%%%%%%%
\section{Multiple excitons} 
%%%%%%%%%%%%%%%%%%%%%%%%%%%%%%%%%%%
%%%%%%%%%%%%%%%%%%%%%%%%%%%%%%%%%%%
\label{multi_exc}

The Hamiltonian given in \eref{Hamil_dressed} is the $N$-body generalisation of the results of \aref{rotwave} with added dipole-dipole interactions. It conserves the exciton-number, represented by the operator \bref{Nexcitons}. Consequently the Hamiltonian has a block-diagonal structure, with each block describing a given number of excitons. For $N=2$ atoms we explicitly have
\begin{eqnarray}
\!\!\!\!\!\!\!\!\!\!\!\!\!\!\!\!\!\!\!\!\!
\label{Hamiltonian_blockstructure}
\hat{H}=
\left[  
\begin{array}{ccc}
M_{0} & 0 & 0 \\
0 & M_{1} & 0 \\
0 & 0 & M_{2} \\
\end{array} 
\right],\:\:
M_{0}=
\left[  
\begin{array}{cccc}
0 & \frac{\Omega_s^*}{2} & \frac{\Omega_s^*}{2} & 0 \\
\frac{\Omega_s}{2} & -\Delta_{s} & 0 &  \frac{\Omega_{s}^*}{2}  \\
\frac{\Omega_s}{2} & 0 & -\Delta_{s} &  \frac{\Omega_{s}^*}{2}  \\
0 & \frac{\Omega_s}{2} & \frac{\Omega_s}{2} & -2 \Delta_{s} \\
\end{array} 
\right] ,\:\:
M_{2}=M_{0}\Big|_{s\rightarrow p}
\end{eqnarray}
\begin{eqnarray}
\label{Hamiltonian_mainblock}
\!\!\!\!\!\!\!\!\!\!\!\!\!\!\!\!\!\!\!\!\!
M_{1}=
\left[  
\begin{array}{cccccccc}
0 & \frac{\Omega_{p}^*}{2} & 0 & 0 & \frac{\Omega_{s}^*}{2} & 0 & 0 & 0  \\
\frac{\Omega_{p}}{2} & -\Delta_{p} & 0 & 0 & 0 & \frac{\Omega_{s}^*}{2} & 0 & 0 \\
0 & 0 & 0 & \frac{\Omega_{s}^{*}}{2} & 0 & 0 &  \frac{\Omega_{p}^*}{2} &  0  \\
0 & 0 & \frac{\Omega_{s}}{2} & -\Delta_{s} & 0 & 0 &  0 & \frac{\Omega_{p}^*}{2}   \\
\frac{\Omega_{s}}{2} & 0 & 0 & 0  & -\Delta_{s} & \frac{\Omega_{p}^*}{2} & 0 & 0  \\
0 & \frac{\Omega_{s}}{2} & 0  & 0 &  \frac{\Omega_{p}}{2} &  -\Delta_{s}  -\Delta_{p} & 0 & U_{12} \\
0 & 0 & \frac{\Omega_{p}}{2} & 0  & 0 &  0 &   -\Delta_{p}  &  \frac{\Omega_{p}^*}{2} \\
0 & 0 & 0 & \frac{\Omega_{p}}{2} & 0  & U_{12} &  \frac{\Omega_{p}}{2} & -\Delta_{s}  -\Delta_{p}  \\
\end{array} 
\right].  
\end{eqnarray}
The bases with respect to which the three blocks are written are $b_{0}=\{\ket{gg},\:\ket{gs},\:\ket{sg},\:\ket{ss} \}$, $b_{1}=\{\ket{gh},\:\ket{gp} ,\:\ket{hg},\:\ket{hs},\:\ket{sh} ,\ket{sp},\:\ket{pg},\:\ket{ps}\}$, $b_{2}=\{\ket{hh},\:\ket{hp},\:\ket{ph},\:\ket{hh} \}$. In \sref{surfaces} we study the nontrivial part, $M_{1}$, of this Hamiltonian in more detail.

%%%%%%%%%%%%%%%%%%%%%%%%%%%%%%%%%%%
%%%%%%%%%%%%%%%%%%%%%%%%%%%%%%%%%%%
\section{Van-Vleck perturbation theory} 
%%%%%%%%%%%%%%%%%%%%%%%%%%%%%%%%%%%
%%%%%%%%%%%%%%%%%%%%%%%%%%%%%%%%%%%
\label{vanvleck}

The basic goal of Van-Vleck perturbation theory was outlined in \sref{dimer_vanVleck}: to find a basis that block-diagonalises the Hamiltonian \bref{Hamil_dressed} to a given order in the perturbation $V$. This appendix supplies the details necessary to understand the origin of the results presented in \sref{dimer_vanVleck} and \ref{sixth_order}. We first partition our unperturbed basis $\ket{\bv{n}}$ into two sets ${\cal P}$ and ${\cal Q}$, the first of which shall span the ``system'' space of interest and the second is its complement in the full Hilbertspace (the ``environment''). The specific basis, $\ket{\bv{n}}$, used for the definition of system and environment is given in \eref{nbodybasis}. This basis is also an eigenbasis of $H_{0}$. Then we can construct projection operators on the system subspace ${P}=\sum_{\phi \in {\cal P}}\ket{\phi}\bra{\phi}$ and its complement ${Q}=\id - {P}$. 
Here we consider the example where ${\cal P}=\{\ket{\pi_{n}},1 \leq n\leq N\}$, thus our system space is the ground-state manifold. We then have  ${P}=\sum_{n}\ket{\pi_{n}}\bra{\pi_{n}}$. 

With respect to these partitions the Hamiltonian matrix, or similarly any other operator, can be divided into four blocks ${P}{H}P$, ${P}{H}{Q}$, $Q{H}{P}$, ${Q}{H}{Q}$. Out of these, we assemble a block diagonal part $H_{D}={P}{H}P + {Q}{H}{Q}$ and a block-off-diagonal part $H_{X}={P}{H}{Q} + {Q}{H}{P}$. Since $H_{0}$ is diagonal in its eigenbasis, we have $H_{D}=H_{0} + V_{D}$,  $H_{X}=V_{X}$. We now aim to find a unitary transformation $T$ that yields a block diagonal Hamiltonian ${\cal H}=T^{-1}HT$ to a given order in~$V$, thus ${\cal H}_{X}=Q{\cal H}P+ P{\cal H}Q=0$. The effective Hamiltonian in the space of interest that we seek, is then given by the block $\sub{H}{eff}=P{\cal H}P$.

Next, we express the unitary transformation operator $T$ as $T=e^G$ with $G=-G^{\dagger}$. We then introduce the condition $G_{D}=0$ and $G_{X}=G$ for the block diagonal and off-diagonal parts of $G$. Other choices are possible, the present one distinguishes the van Vleck procedure from other related schemes \cite{shavitt:vanVleck}. For all operators in the problem, we write a series expansion in orders of the perturbation $V$, most notably ${\cal H}=H_{0} + \sum_{n=1}^{\infty}W^{(n)}$, $T= \sum_{n=1}^{\infty}T^{(n)}$, $G= \sum_{n=1}^{\infty}G^{(n)}$. 

As outlined in \cite{shavitt:vanVleck}, it is then possible to obtain recursive relations for $\comm{H_{0}}{G^{(n)}}$\footnote{
A crucial step is to rewrite ${\cal H}=e^{-G}H e^{G}$ as series of increasingly deeper nested commutators using the Baker-Campbell-Hausdorff formula. By segregating expansion orders in $V$, the equation is then turned into a recursive set of equations for the $\comm{H_{0}}{G^{(n)}}$. An elegant method to perform this cumbersome routine is outlined in \cite{shavitt:vanVleck}, to which we refer for further details.
}.
One extracts the full matrix $G^{(n)}$ from $\comm{H_{0}}{G^{(n)}}$ as follows. Let us introduce the notation $\ket{i}$ with latin indices to denote an eigenvector of $H_{0}$ within ${\cal Q}$ and $\ket{\alpha}$ with greek indices for one within ${\cal P}$. Since we demand $G_{D}=0$ we only require matrix elements like $\bra{i}G^{(n)}\ket{\alpha}$ or $\bra{\alpha}G^{(n)}\ket{i}$ in order to know the whole matrix form of $G^{(n)}$. Now we define the resolvent operator. 
\begin{eqnarray}
R_{\alpha}^{(0)}=\sum_{i}\frac{\ket{i}\bra{i}}{\epsilon_{\alpha} - \epsilon_{i}},
\label{resolvent}
\end{eqnarray}
which fulfills:
\begin{eqnarray}
O\ket{\alpha}=-R_{\alpha}^{(0)}\comm{H_{0}}{O} \ket{\alpha}.
\label{resolvent_eqn}
\end{eqnarray}
for any operator $O$. Substituting $O=G^{(n)}$ in \eref{resolvent_eqn} and multiplying from the left with $\bra{i}$ we obtain $\bra{i}G^{(n)}\ket{\alpha}$ and also $\bra{\alpha}G^{(n)}\ket{i}=-\bra{i}G^{(n)}\ket{\alpha}^*$.

From knowledge of $G$ one can infer the expansion orders $W^{(n)}$ through commutation relations. A series expansion of $T=e^G$ to the required order finally yields the perturbed eigenstates
via 
\begin{eqnarray}
T\ket{\alpha}=\sum_{t} \ket{t}\bra{t}T\ket{\alpha}.
\label{perturbed_states}
\end{eqnarray}
We refer to \rref{shavitt:vanVleck} for all further technical details.

%%%%%%%%%%%%%%%%%%%%%%%%%%%%%%%%%%%
\subsection{Higher orders and atom numbers} 
%%%%%%%%%%%%%%%%%%%%%%%%%%%%%%%%%%%
\label{sixth_order}

The effective Hamiltonian of the dimer in the ground-state manifold spanned by $\ket{\tilde{\pi}_{1,2}}$ is given in \sref{dimer_vanVleck} up to fourth order, with odd orders vanishing. Here we display the corresponding expression of sixth order, $H_{C}^{(6)}$, where we use the definition $\sub{H}{eff}^{(m)}=PW^{(m)}P$.
We can write
\begin{eqnarray}
\sub{H}{eff}^{(6)}=
\left(
\begin{array}{ccc}
W^{(6)} (R)& \tilde{U}_{12}^{(6)}(R)   \\
\tilde{U}_{21}^{(6)} (R) & W^{(6)} (R)  \\
\end{array}
\right),
\label{ER6}
\end{eqnarray}
with
\begin{eqnarray}
W^{(6)}(R) &=\Bigg\{
2(\alpha_{s}^6 \Delta_s+  \alpha_{p}^6\Delta_p) + (\alpha_{s}^2 \Delta_s+  \alpha_{p}^2 \Delta_p)\sub{\bar{U}}{12}^2\times
\CR
&\Bigg[
-4(\alpha_{s}^4 +  \alpha_{p}^4) - \alpha_{s}^2 \alpha_{p}^2\frac{\Delta_s^2 -4 \Delta_s \Delta_p + \Delta_p^2}{\Delta_s \Delta_p}
\CR
&+\left(
2(\alpha_{s}^4 +  \alpha_{p}^4) + \alpha_{s}^2 \alpha_{p}^2\frac{(\Delta_s + \Delta_p)^2}{\Delta_s \Delta_p}
 \right)\sub{\bar{U}}{12}^2
\Bigg]
\Bigg\}
/
(\sub{\bar{U}}{12}^2 - 1)^2,
\label{Wpotentials6}
\\
\tilde{U}_{12}^{(6)}(R) &=  \alpha_{s}^2 \alpha_{p}^2 \sub{U}{12}\Bigg[ \frac{  \alpha_{s}^2 +  \alpha_{p}^2 }{\sub{\bar{U}}{12}^2 - 1}
\CR
&+\frac{-2 (\alpha_{s}^2 +  \alpha_{p}^2 ) + 2 \sub{\bar{U}}{12}^2  \frac{(\Delta_{s}^2 + \Delta_{p}^2)}{(\Delta_{s} + \Delta_{p})} 
\left( \frac{\alpha_{p}^2}{\Delta_{s} }  +  \frac{\alpha_{s}^2}{\Delta_{p} } \right)  }
{(\sub{\bar{U}}{12}^2 - 1)^2}
\Bigg].
\label{Vpotentials6}
\end{eqnarray}
Recall that$\bar{\Delta}=\Delta_{s} + \Delta_{p}$ and $\sub{\bar{U}}{12}=\sub{U}{12}/\bar{\Delta}$.

Finally, for both cases shown in \sref{applications}, the effective Hamiltonian of \sref{dimer_vanVleck} has to be adjusted for $N>2$. In terms of the basis $\ket{\tilde{\pi}_{n}}$ we obtain: 
\begin{eqnarray}
\sub{H}{eff,ij}&=(E_{2} + E_{4})\delta_{ij} + \tilde{U}_{ij}(R),
\label{HeffN}
\\
E_{2} &=  (N-1)\alpha_{s}^2 \Delta_{s}+  \alpha_{p}^2\Delta_{p}, 
\\
E_{4} &= -[(N-1)\alpha_{s}^4 \Delta_{s}+  \alpha_{p}^4\Delta_{p} +   (N-1)\alpha_{s}^2 \alpha_{p}^2(\Delta_{s} + \Delta_{p})],
\\
\tilde{U}_{jj}(R) &= \alpha_{s}^2 \alpha_{p}^2 \left( \sum_{k\neq j} \frac{1}{1 -  \sub{\bar{U}}{kj}^2 }  \right)(\Delta_{s} + \Delta_{p}) \delta_{ij} 
\CR
&+  \alpha_{s}^2 \alpha_{p}^2 \frac{ \sub{U}{ij}  }{1 - \sub{\bar{U}}{ij}^2} (1-\delta_{ij}).
\end{eqnarray}
%

%%%%%%%%%%%%%%%%%%%%%%%%%%%%%%%%%%%
%%%%%%%%%%%%%%%%%%%%%%%%%%%%%%%%%%%

\end{document}